\documentclass[fleqn,10pt]{wlscirep}
\usepackage[utf8]{inputenc}
\usepackage[T1]{fontenc}
\usepackage{hyperref}
\hypersetup{
  colorlinks = true,
  citecolor = blue,
  urlcolor = blue
}

\usepackage{braket}
\usepackage{graphicx}
\usepackage[mathscr]{euscript}
\usepackage{amsmath}
\usepackage{amsfonts}
\usepackage{bm}
\usepackage{array}
\usepackage{float}

\newcolumntype{x}[1]{>{\hfil$\displaystyle} p{#1} <{$\hfil}}
\newcommand{{\bfL}}{\mbox{\boldmath$L$\unboldmath}}
\newcommand{{\bfl}}{\mbox{\boldmath$l$\unboldmath}}
\newcommand{{\bftheta}}{\mbox{\boldmath$\theta$\unboldmath}}
\newcommand{{\bfq}}{\mbox{\boldmath$q$\unboldmath}}
\newcommand{{\bfg}}{\mbox{\boldmath$g$\unboldmath}}
\newcommand{{\bfdelta}}{\mbox{\boldmath$\delta$\unboldmath}}
\newcommand{{\bfeta}}{\mbox{\boldmath$\eta$\unboldmath}}
\newcommand{{\bfmu}}{\mbox{\boldmath$\mu$\unboldmath}}
\newcommand{{\bfbeta}}{\mbox{\boldmath$\beta$\unboldmath}}
\newcommand{{\bfphi}}{\mbox{\boldmath$\phi$\unboldmath}}
\newcommand{{\bfrho}}{\mbox{\boldmath$\rho$\unboldmath}}
\newcommand{{\bfcalB}}{\mbox{\boldmath$\cal B$\unboldmath}}
\newcommand{{\bfcalE}}{\mbox{\boldmath$\cal E$\unboldmath}}
\newcommand{{\bfC}}{\mbox{\boldmath$\cal C$\unboldmath}}
\newcommand{{\bfcalJ}}{\mbox{\boldmath$\cal J$\unboldmath}}
\newcommand{{\bfcalO}}{\mbox{\boldmath$\cal O$\unboldmath}}
\newcommand{{\bfA}}{\mbox{\boldmath$\emph{A}$\unboldmath}}
\newcommand{{\bfB}}{\mbox{\boldmath$\emph{B}$\unboldmath}}

\newcommand{{\bfcalR}}{\mbox{\boldmath$\cal R$\unboldmath}}

\def\v#1{{\bf#1}}

\DeclareMathAlphabet\mathbfcal{OMS}{cmsy}{b}{n}

\title{\LARGE{Topology, nonlocality and duality in classical electrodynamics}}
\author[*]{\normalsize{Jos\'e A. Heras}}
\affil[*]{Instituto de Geof\'isica, Universidad Nacional Aut\'onoma de M\'exico, Ciudad de M\'exico 04510, M\'exico }

\author[**]{\normalsize{ Ricardo Heras}}
\affil[**]{Department of Physics and Astronomy, University College London, London WC1E 6BT, UK
 }

\begin{abstract}
We have recently \cite{1} argued that classical electrodynamics can predict nonlocal effects by showing an example of a topological and nonlocal electromagnetic angular momentum. In this paper we discuss the dual of this angular momentum which is also topological and nonlocal. We then unify both angular momenta by means of the electromagnetic angular momentum arising  in the configuration formed by a dyon encircling an infinitely-long dual solenoid enclosing uniform electric and magnetic fluxes and show that this electromagnetic angular momentum is topological because it depends on a winding number, is nonlocal because the electric and magnetic fields of this dual solenoid act on the dyon in regions for which these fields are excluded and is invariant under electromagnetic duality transformations. We explicitly verify that this  duality-invariant electromagnetic angular momentum is insensitive to the radiative effects of the Li\'enard-Wiechert fields of the encircling dyon. We also show how duality symmetry of this angular momentum  suggests different physical interpretations for the corresponding angular momenta that it unifies.
\end{abstract}
\begin{document}

\flushbottom
\maketitle

\thispagestyle{empty}
\section{Introduction}
In a recent paper \cite{1} we have pointed out that classical electrodynamics can predict nonlocal effects when considering
electromagnetic configurations lying in non-simply connected regions. We have shown that the configuration formed by an electric charge $q$ encircling an infinitely-long magnetic solenoid enclosing a uniform magnetic flux $\Phi_m$ accumulates the electromagnetic angular momentum
\begin{equation}
\bfL_q=\frac{n q \Phi_m}{2\pi c}\hat{\v z},
\end{equation}
where the integer $n$ specifies the number of times the electric charge encircles the magnetic solenoid. The electric charge moves in a non-simply connected region where there is no magnetic field and therefore there is no Lorentz force acting on the electric charge but there is a non-zero vector potential. The electromagnetic angular  momentum $\bfL_q$ is topological because it depends on the winding number $n$ and is nonlocal because the magnetic field of the solenoid acts on the electric charge in regions for which this field is excluded. We have argued that the magnitude of (1) can be considered as the classical counterpart of the Aharonov-Bohm (AB) phase \cite{2}: $\delta_{\rm AB}=q\Phi_m/(\hbar c)$, being both quantities connected by the linear relation \cite{3,4}: $\delta_{\rm AB}=2\pi L_q/\hbar.$

In this paper we discuss two other examples of nonlocal electromagnetic angular momenta arising in non-simply connected regions. The first example is the electromagnetic angular momentum
\begin{equation}
\bfL_g=-\frac{n g \Phi_e}{2\pi c}\hat{\v z},
\end{equation}
of the dual configuration formed by a magnetic charge $g$ encircling an infinitely-long electric solenoid enclosing a uniform electric flux $\Phi_e$. The electromagnetic angular  momentum $\bfL_g$ is the dual of the electromagnetic angular  momentum $\bfL_q$, i.e., the former can be obtained from the latter by simply making the dual changes $q\to g$ and $\Phi_m\to -\Phi_e$. We show that $\bfL_g$ is also topological and nonlocal and suggest that its $z-$component can be considered as the classical counterpart of the dual of the Aharonov-Bohm (DAB) phase
\cite{5,6}: $\delta_{\rm DAB}=-gn\Phi_e/(\hbar c)$, being both quantities connected by the linear relation $\delta_{\rm DAB}=2\pi L_g/\hbar.$

The second example is the duality-invariant electromagnetic angular momentum
\begin{equation}
\bfL_{qg}= \frac{n(q\Phi_m-g\Phi_e)}{2\pi c}\hat{\v z},
\end{equation}
of the configuration formed by a dyon \cite{7}, a particle possessing both an electric charge $q$ and a magnetic charge $g$,  encircling an idealised infinitely-long dual solenoid enclosing both a uniform electric flux $\Phi_e$ and a uniform magnetic flux $\Phi_m$. We show that $\bfL_{qg}$ is topological because it depends on the winding number $n$, is nonlocal because the electric and magnetic fields of this dual solenoid act on the dyon in regions for which these fields are excluded and is invariant under two sets of independent  electromagnetic duality transformations: the set formed by the transformations $q\to g$ and $g\to -q,$ and the set formed by the transformations $\Phi_m\to -\Phi_e$ and $\Phi_e\to \Phi_m$.
More in general, we show that $\bfL_{qg}$ is invariant under a $U(1)$ duality transformation group having the angle $\theta$ as the transformation parameter. We then argue that $\bfL_{qg}$ provides a unified description of $\bfL_q$ and $\bfL_g,$ which means that $\bfL_{qg}$ becomes $\bfL_q$  for certain values of $\theta$ and $\bfL_g$ for other values of $\theta$. This duality symmetry is not only interesting from mathematical point of view but is also interesting from a physical point of view since it allows us to give two different physical interpretations for both $\bfL_q$ and $\bfL_g$. We then claim
that the $z$-component of (3) can be considered as the classical counterpart of the duality-invariant quantum phase introduced in
\cite{8}: $\delta_{\rm D}=(q\Phi_m-g\Phi_e)/(\hbar c)$, being both quantities connected by the linear relation $\delta_{\rm D}=2\pi L_{qg}/\hbar.$
We emphasize the topological nature of $\bfL_{qg}$ by noting that it is independent from the dynamics of the dyon and stress this independence by demonstrating that $\bfL_{qg}$ insensitive to the radiative effects of the Li\'enard-Wiechert fields of the encircling dyon.

The fact that no magnetic charges have been observed in nature has led to the common idea that electromagnetic duality is an accidental symmetry of some quantum field theories rather than a fundamental feature of classical electrodynamics \cite{9,10,11}. Duality symmetry of point electric and magnetic charges is manifested in the fact that the external electric and magnetic fields of these charges exhibit the same form. These fields are not defined inside the point charges and therefore duality symmetry of point charges can be thought of as an external symmetry of these charges. On the other hand, classical electrodynamics predicts an exact similarity between the external field lines of an electric dipole and those due to a magnetic dipole, i.e., the external fields of point electric and magnetic dipoles exhibit the same form. This dipole similarity turns out to be  a clear manifestation of duality symmetry, which is observed in the static and dynamical regimes \cite{12}. The difference now is that the electric and magnetic dipole fields are defined inside the point dipoles themselves through delta terms of different form ---therefore the dipole duality symmetry is broken inside the point dipoles. The dipole duality symmetry may be represented by the $U(1)$ duality transformations \cite{12}: $\v E'+i\v B'={\rm e}^{i\theta}\big(\v E+i\v B\big)$ and $\v d'+i\bfmu'={\rm e}^{i\theta}\big(\v d+i\bfmu\big)$, where the external electric and magnetic fields are produced by the electric and magnetic dipole moments. Why has this duality symmetry manifested itself in the electrodynamics of dipoles but not in the electrodynamics of charges? We do not know the answer but the fact that this symmetry exists in electric and magnetic dipoles encourages us to continue looking for monopoles. In any case, the search for magnetic monopoles continues to be of considerable interest \cite{13,14}, mainly motivated by the fact that the elusive magnetic charges would allow us to explain the observed quantisation of the electric charge \cite{15,16}. On the other hand, the MoEDAL collaboration has recently made the first experimental search for dyons \cite{17} which would indirectly prove the existence of magnetic charges. Our study on the electromagnetic angular momentum of the dyon-solenoid configuration has been motivated to some extent by this recent interest in the search for dyons.

\section {Monopole-solenoid configuration}
In this section and in the two subsequent sections we closely follow the discussion given in Ref. \cite{1}
by making appropriate dual changes and putting more emphasis on relevant results than on formal details which can be found in Ref. \cite{1}.

The monopole-solenoid configuration consists of a particle of magnetic charge $g$ and mass $m$ which is continuously moving around an infinitely-long electric solenoid of radius $R$ which encloses a uniform electric flux $\Phi_e$. The $z$-axis is chosen to be the axis of this electric solenoid (See Fig.~1). Cylindrical coordinates $(\rho,\theta,\phi)$ are adopted. The corresponding magnetic current density reads
\begin{equation}
\v J_m = -\frac{c\,\Phi_e\delta(\rho-R)}{4\pi^2 R^2}\,\hat{\!\bfphi},
\end{equation}
where $\delta(\rho-R)$ is the Dirac delta function and $\Phi_e = \pi R^2 E$ is the electric flux through the electric solenoid with $E$ being the magnitude of the uniform electric field inside this electric solenoid. The current $\v J_m$ is a steady current: $\nabla\cdot \v J_m=0$ and its electric field satisfies the equations
\begin{equation}
\nabla \cdot \v E =0,\quad \nabla \times\v E= \frac{\Phi_e\delta(\rho-R)}{\pi R^2}\,\hat{\!\bfphi},
\end{equation}
whose solution is given by the electric field
\begin{equation}
\v E=\frac{\Phi_e \Theta(R-\rho)}{\pi R^2}\hat{\v z},
\end{equation}
which is  confined in the electric solenoid. Here $\Theta(\rho-R)$ is the Heaviside step function.
From (6) it follows that $\v E_{\rm out}=0$ and  $\v E_{\rm in}=\Phi_e\hat{\v z}/(\pi R^2),$
where $\v E_{\rm out}$ and $\v E_{\rm in}$ are respectively the electric field  outside $(\rho>R)$ and inside $(\rho<R)$ the electric solenoid. From the first equation in (5) we infer $\v E = -\nabla \times \v C$ where $\v C$ is the corresponding electric vector potential. Using this relation in the second equation in (5) and adopting the Coulomb gauge $\nabla \cdot \v C=0$, we obtain the Poisson equation
\begin{equation}
\nabla^2 \v C = \frac{\Phi_e\delta(\rho-R)}{\pi R^2}\,\hat{\!\bfphi},
\end{equation}
whose solution reads
\begin{equation}
\v C= -\frac{\Phi_e}{2 \pi}\bigg(\frac{\Theta(\rho-R)}{\rho}+\frac{\rho\,\Theta(R-\rho)}{R^2}\bigg)\,\hat{\!\bfphi}.
\end{equation}
The proof that the Laplacian of (8) yields  (7) is entirely similar to that given  in Appendix A of Ref. \cite{1} with the changes $\v A\to \v C $ and $\Phi_m\to -\Phi_e$. The potential $\v C$ is not defined at $\rho=R$ but one can regularise it to obtain $\v C(R)= -\Phi_e\,\hat{\!\bfphi}/(2\pi R)$ which indicates that $\v C$ is continuous at $\rho=R.$ Equation (8) satisfies the Coulomb gauge: $\nabla \cdot \v C= (1/\rho)(\partial C_{\phi}/\partial \phi)=0.$ The proof that minus the curl of (8) yields the electric field in (6) is entirely similar to that given in Appendix B of Ref. \cite{1} with the dual changes $\v A\to \v C $ and $\Phi_m\to -\Phi_e$. From (8) it follows that
\begin{equation}
\v C_{\rm out}= -\frac{\Phi_e}{2 \pi\rho}\,\hat{\!\bfphi},\quad \v C_{\rm in}= -\frac{\rho\Phi_e}{2 \pi R^2}\,\hat{\!\bfphi},
\end{equation}
where $\v C_{\rm out}$ is the electric vector potential outside the electric solenoid $(\rho>R)$, which is  connected with its electric field $\v E_{\rm out}=-\nabla \times \v C_{\rm out}=0$ and $\v C_{\rm in}$ is the electric vector potential inside the electric solenoid $(\rho<R)$, which is  connected with its electric field $\v E_{\rm in}=-\nabla \times \v C_{\rm in}=\Phi_e\hat{\v z}/(\pi R^2).$
\begin{figure}
	\centering
	\includegraphics[scale=0.35]{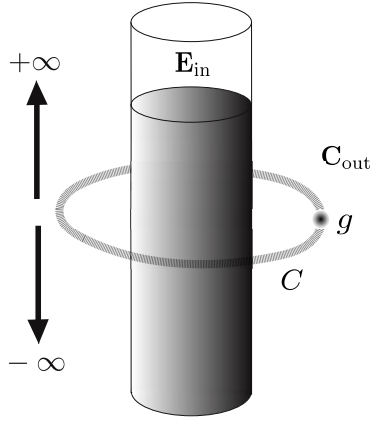}
	\caption{\normalsize{Monopole-solenoid configuration. A magnetic charge moving in the $x$-$y$ plane along the path $C$ which encircles an infinitely-long solenoid enclosing a uniform electric flux.}}
	\label{Fig1}
\end{figure}
The potential $\v C_{\rm out}$  is a pure gauge potential: $\v C_{\rm out}= \nabla \chi_e$ with $\chi_e=-\Phi_e \phi/(2\pi)$ being
a multi-valued function that satisfies $\nabla^2\chi_e=0.$  The Lorentz force reads $\mathbf{F} = -g \dot{\v x}\times \v E/c$, where $\dot{\v x}=d\v x/dt$ is the velocity of the magnetic charge and $\v x$ is its position. Inserting the electric field (6) and $\dot{\v x} = \dot{\rho}\,\hat{\!\bfrho} + \rho \dot{\phi}\,\hat{\!\bfphi}+ \dot{z}\hat{\v z}$ in the Lorentz force, we obtain
\begin{equation}
\v F = -\frac{g\Phi_e \Theta(R-\rho)}{\pi R^2}(\rho \dot{\phi}\,\hat{\!\bfrho}-\dot{\rho}\,\hat{\!\bfphi}).
\end{equation}
In the monopole-solenoid configuration the moving magnetic charge is outside the solenoid $(\rho>R)$, and therefore $\Theta=0$ which implies
$\v F = 0.$ In short: the Lorentz force vanishes because the electric field is zero in the region where the magnetic charge is moving.
%\end{document}
\section{Topology and nonlocality of the circulation of $\v C_{\rm out}$}
In Sec.~3 of Ref. \cite{1} we have demonstrated that the Cauchy's integral formula for an analytic function implies the particular relation
\begin{eqnarray}
\oint_C \bigg(\frac{K\,\hat{\!\bfphi}}{2\pi\rho} \bigg)\cdot d \v x=\left\{\begin{array}{@{}l@{\quad}l}
       nK\,& \mbox{if $C$ encloses $\rho=0$} \\[\jot]
      \,0 & \mbox{otherwise}
    \end{array}\right.
\end{eqnarray}
where $K$ is a constant and $n$ is the winding number of the curve $C$ which gives the number of times the curve $C$ encircles the singularity $\rho=0$. Let us now apply (11) to the infinitely-long electric solenoid enclosing the uniform electric flux $\Phi_e$. We make the identification $K=-\Phi_e$ in (11) and obtain the relation $K\,\hat{\!\bfphi}/(2\pi\rho)=-\Phi_e\,\hat{\!\bfphi}/(2 \pi \rho)=\v C_{\rm out}(\v x)$. Since the solenoid encloses the singularity at $\rho=0$ then it follows
\begin{eqnarray}
\oint_C\!\v C_{\rm out}\!\cdot \!d \v x=\left\{\begin{array}{@{}l@{\quad}l}
       -n\Phi_e & \mbox{if $C$ encloses the electric solenoid} \\[\jot]
      \,0 & \mbox{otherwise}
    \end{array}\right.
\end{eqnarray}
which states that if the curve $C$ encircles $-n$ times the electric solenoid then $\oint_C\v C_{\rm out}\cdot d \v x$ accumulates $-n$ times the electric flux $\Phi_e$. Since $-n\Phi_e$ is a constant quantity then $\oint_C\v C_{\rm out}\cdot d \v x$ is insensitive to the form of the curve $C$ and thereby to the dynamics that we could associate to it. If we consider $C_1,C_2...C_k$ different curves which enclose $-n$ times  the electric solenoid then (12) implies
\begin{eqnarray}
\oint_{C_1}\v C_{\rm out} \cdot d \v x = \oint_{C_2}\v C_{\rm out}\cdot \!d \v x=...=\oint_{C_k}\v C_{\rm out} \cdot d \v x.
\end{eqnarray}
The curves $C_1,C_2...C_k$ are homotopically equivalent and thus we could not distinguish if  $-n\Phi_e$ is connected with the circulation of $\v C_{\rm out}$ along $C_1$ or along $C_2$ or along  $C_k$.
\begin{figure}
	\centering
	\includegraphics[scale=0.6]{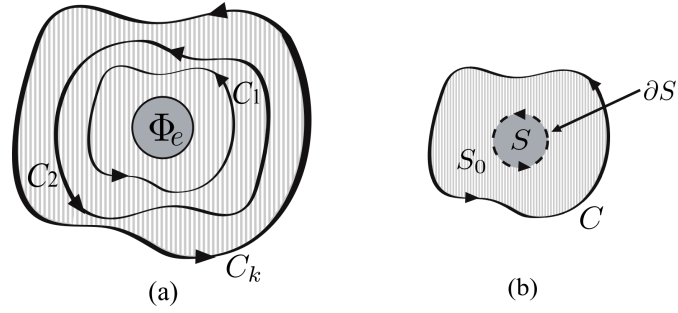}
	\caption{\normalsize{(a) The circulation of $\v C_{\rm out}$ in a closed path around  the electric solenoid is insensitive to the form of the path. (b) The circulation $\v C_{\rm out}$ is taken along a path $C$ greater than the boundary $\partial S$ of the surface $S$ of the electric solenoid. Since $\oint_{C>\partial S}\v C_{\rm out}\cdot d \v x=-\int_S \v E_{\rm in} \cdot d \v S$ holds then the circulation of $\v C_{\rm out}$ is spatially delocalised from the surface of the solenoid where the flux of the electric field is localised.}}
	\label{Fig2}
\end{figure}
Following Ref. \cite{1} we can show that the Stokes theorem applied to the monopole-solenoid configuration takes the form,
\begin{equation}
\oint_{C>\partial S}\v C_{\rm out}\cdot d \v x =\int_{S}\nabla \times \v C_{\rm in}\cdot d\v S,
\end{equation}
 Considering $\nabla \times \v C_{\rm in}=-\v E_{\rm in}$ with $\v E_{\rm in}$ being the constant electric field inside the electric solenoid, (14) becomes
\begin{equation}
\oint_{C>\partial S}\v C_{\rm out}\cdot d \v x = -\int_{S}\v E_{\rm in}\cdot d\v S,
\end{equation}
which admits a nonlocal interpretation: while the left-hand side is defined outside the electric solenoid, the right-hand side is defined inside this solenoid, i.e., $\oint_{C>\partial S}\v C_{\rm out}\cdot d \v x$ and $\int_{S}\v E_{\rm in}\cdot d\v S,$ are nonlocally connected. Equations (13) and (15) imply the relation
\begin{equation}
\oint_{C_k>\partial S}\! \v C_{\rm out}\cdot d \v x=...=\oint_{C_2>\partial S} \!\v C_{\rm out}\cdot d \v x=\oint_{C_1>\partial S} \!\v C_{\rm out}\cdot d \v x=-\int_{S} \v E_{\rm in} \cdot d \v S,
\end{equation}
according to which we cannot distinguish if $-n\Phi_e$ is connected with the circulation of $\v C_{\rm out}$ along $C_1>\partial S$ or along $C_2>\partial S$ or along  $C_k>\partial S$ (see Fig.~2). The circulations in (16) are spatially delocalised with respect to the electric flux. We then conclude that the potential  $\v C_{\rm out}$ is ambiguous due to its gauge-dependence and its circulation $\oint_{C}\v C_{\rm out}\cdot d \v x$ is also ambiguous due to its spatial delocalisation (indistinguishability of $C$).
\section{Electromagnetic angular momentum of the monopole-solenoid configuration}
The following theorem is the dual of the theorem formulated in Sec.~4 of Ref. \cite{1}.
\vskip 3pt
\noindent \emph{Dual Decomposition Theorem.} Let $\bfcalB(\v x,t)$ be a time-dependent magnetic field and $\varrho_m(\v x,t)\!=\!\nabla\cdot \bfcalB/(4\pi)$ its associated magnetic charge density. Let $\v E(\v x)$ be a time-independent electric field and ${\v C}(\v x)$ its associated vector potential $\nabla\times \v C=-\v E$. The fields $\bfcalB$ and ${\v E}$ are independent fields produced by different sources. The Poynting formula for the electromagnetic angular momentum in the volume $V$ originated by the interaction between the fields $\bfcalB$ and $\v E$ is given by
\begin{equation}
\textbf{\bfL}_{\rm P}=\frac{1}{4\pi c}\int_{V}\v x \times (\v E\times \bfcalB)\,d^3x,
\end{equation}
which can be decomposed as
\begin{equation}
\textbf{\bfL}_{\rm P}=\textbf{\bfL}_{\rm M}+\textbf{\bfL}_{\rm R}+\textbf{\bfL}_{\rm G}+\textbf{\bfL}_{\rm S},
\end{equation}
where
\begin{eqnarray}
\textbf{\bfL}_{\rm M}\!\!\!\!\!&=&\!\!\!\!\frac{1}{c}\int_{V}\v x \times (\varrho_m\v C)\,d^3x, \\
\textbf{\bfL}_{\rm R}\!\!\!\!\!&=&\!\!\!\!\!\frac{1}{4\pi c}\int_{V}\v x \times \big[(\nabla\times\bfcalB)\times \v C\big]\,d^3x,\\
\textbf{\bfL}_{\rm G}\!\!\!\!\!&=&\!\!\!\!\!\frac{1}{4\pi c}\int_{V}\v x \times (\bfcalB\nabla\cdot \v C)\,d^3x,\\
\textbf{\bfL}_{\rm S}\!\!\!\!\!&=&\!\!\!\!\!\frac{1}{4\pi c}\oint_{S}\v x \times\big[\hat{\textbf{n}}(\bfcalB\cdot
\textbf{C})-\textbf{C}(\hat{\v n} \cdot\bfcalB)-\bfcalB(\hat{\textbf{n}}\cdot \textbf{C})\big]\,d{S}.
\end{eqnarray}
Here $S$ is the surface of the volume $V$. The proof and interpretation of this theorem is entirely similar to that formulated Ref. \cite{1} with the dual changes $\bfcalE\to \bfcalB, \v B\to -\v E$ and $\v A\to \v C $. The term $\textbf{\bfL}_{\rm R}$ in (20) deals with possible radiative effects. The term $\textbf{\bfL}_{\rm G}$ in (21) is determined by the adopted gauge for the potential $\v C$. The term $\textbf{\bfL}_{\rm S}$ in (22) is a surface term.

We now proceed to apply (18) to the monopole-solenoid configuration where the volume $V$ covers all space except the volume of the electric solenoid. It is clear $\textbf{\bfL}_{\rm P}=0$ because $\textbf{E}=0$ outside the electric solenoid and therefore (18) becomes
$0\!=\!\textbf{\bfL}_{\rm M}+\textbf{\bfL}_{\rm R}+\textbf{\bfL}_{\rm G}+\textbf{\bfL}_{\rm S}$. At the end of the Appendix E we show that $\textbf{\bfL}_{\rm R}=0$ indicating that the electromagnetic angular momentum of the monopole-solenoid configuration is insensitive to the Li\'enard-Wiechert fields of the encircling magnetic charge. We have also  $\textbf{\bfL}_{\rm G}=0$ because of the adopted Coulomb gauge $\nabla\cdot \v C_{\rm out}=0$. Therefore (18) reduces to $0=\textbf{\bfL}_{\rm M}+\textbf{\bfL}_{\rm S}$ and thus there exists $\textbf{\bfL}_{\rm M}$ in the volumetric space  where the magnetic  charge is moving around the electric solenoid and there exists $\textbf{\bfL}_{\rm S}$ on the surface of this volumetric space which lies at infinity. We will see that the piece $\textbf{\bfL}_{\rm M}$ is related to the flux of the electric field of the solenoid  and therefore $\textbf{\bfL}_{\rm S}$ may be interpreted as an electromagnetic angular momentum originated by the return flux of the electric  field of an infinitely-long electric solenoid. When considering $0=\textbf{\bfL}_{\rm M}+\textbf{\bfL}_{\rm S}$ we should always have in mind that $\textbf{\bfL}_{\rm M}$ and $\textbf{\bfL}_{\rm S}$ are defined in different spatial regions.

In order to obtain the electromagnetic angular  momentum ${\bfL}_g$ of the monopole-solenoid configuration we apply (19) by assuming that the charge $g$ is localised in the $x$-$y$ plane and that its corresponding position vector is given by $\v x_g(t)=\{\rho_g(t)\cos\phi_g(t),\rho_g(t)\sin\phi_g(t),0\}=\rho_g(t)\,\hat{\!\bfrho}$. The associated magnetic charge density takes the form
$\varrho_m(\v x',t)=(g/\rho') \delta\{\rho' - \rho_g(t)\}\delta\{\phi' - \phi_g(t)\}\delta\{z'\}$ and the electric vector potential is given by $\v C_{\rm out}(\v x')=-\Phi_e\, \hat{\!\bfphi}/(2\pi\rho')$. A generic point reads $\v x'\! = \!\rho'\hat{\bfrho} \!+\!z'\hat{\v z}$. Using these ingredients and integrating the right-hand side of (19), we obtain
\begin{equation}
\frac{1}{c}\int_{V}\varrho_m(\v x',t)\, \v x' \times \v C_{\rm out}(\v x')\,d^3x'=-\frac{n g \Phi_e}{2\pi c}\hat{\v z},
\end{equation}
where the winding number $n$ gives the number of times the magnetic charge travels its closed path around the electric solenoid. In the derivation of (23), we have used  the following results: $\int^{\infty}_{-\infty}\delta(z')dz'=1,\int^{\infty}_{-\infty}z'\delta(z')dz'=0,
\int^{\infty}_{R}\delta\{\rho'-\rho_g(t)\}\,d\rho'=\Theta\{\rho_g(t) - R\}=1,$
(because $\rho_g(t)>R$ outside the electric solenoid) and the relation $\oint_{C} \delta\{\phi' - \phi_g(t)\}\,d\phi'=n$
which is demonstrated in Appendix F of Ref. \cite{1}. Therefore (23) represents the accumulated electromagnetic angular momentum of the monopole-solenoid configuration
\begin{equation}
\bfL_g=-\frac{n g \Phi_e}{2\pi c}\hat{\v z}.
\end{equation}
Using (12) we can also write
\begin{equation}\,
\bfL_g=\frac{g\hat{\v z}}{2\pi c}\oint_{C} \v C_{\rm out}\cdot d \v x,
\end{equation}
which is gauge invariant on account of the gauge invariance of the circulation of the potential $\v C_{\rm out},$ i.e.,
$\oint_C \v C_{\rm out}' \cdot d \v x= \oint_C \v C_{\rm out} \cdot d \v x+ \oint_C \nabla\Lambda \cdot d \v x= \oint_C \v C_{\rm out} \cdot d \v x,$ where $\Lambda$ is a  single-valued gauge function. According to (25) the electromagnetic angular  momentum $\bfL_{g}$ depends on the circulation of the potential $\v C_{\rm out}$.  Using (15) we can see that (25) can be expressed as
\begin{equation}
\bfL_g= -\frac{g\hat{\v z}}{2\pi c}\int_{S}\v E_{\rm in}\cdot d\v S,
\end{equation}
which states that the electromagnetic angular momentum $\bfL_{g}$ calculated outside the electric solenoid depends on the flux of the electric field inside this solenoid. Evidently, (25) and (26) suggest different interpretations of the physical origin of the electromagnetic angular momentum $\bfL_g$. Using these two equations we obtain the relation
\begin{equation}
\frac{g\hat{\v z}}{2\pi c}\oint_{C} \v C_{\rm out}\cdot d \v x=\bfL_g= -\frac{g\hat{\v z}}{2\pi c}\int_{S}\v E_{\rm in}\cdot d\v S.
\end{equation}
The first interpretation, which we will call the ${\bm C}$-explanation, is supported by the first equality in (27). This states that
$\v C_{\rm out}$ locally acts through its circulation on the magnetic charge originating $\bfL_{g}$. In other words: $\v C_{\rm out}$ exists in each point of the trajectory of $g$ and therefore $\v C_{\rm out}$ locally acts on $g$ producing $\bfL_g$. The second interpretation, which we will call the ${\bm E}$-explanation, is supported by the second equality in (27), which states that  $\v E_{\rm in}$ nonlocally acts through its electric flux on the magnetic charge originating $\bfL_g$. Since both $g$ and $\v E_{\rm in}$ lie in different regions their interaction is nonlocal. In other words: $\v E_{\rm in}$ exists inside the electric solenoid and not in each point of the trajectory of $g$ outside this solenoid and therefore $\v E_{\rm in}$ nonlocally acts on $g$ producing $\bfL_g$. We find unsatisfactory the $\bfC$-explanation for the following arguments:
	the potential $\v C_{\rm out}$ is gauge-dependent and therefore it cannot represent a physical quantity. On the other hand, the relation (16) implies
\begin{equation}
\frac{g\hat{\v z}}{2\pi c}\oint_{C_k>\partial S}\! \v C_{\rm out}\cdot d \v x=...=\frac{g\hat{\v z}}{2\pi c}\oint_{C_2>\partial S} \!\v C_{\rm out}\cdot d \v x=\!\frac{g\hat{\v z}}{2\pi c}\oint_{C_1>\partial S} \!\v C_{\rm out}\cdot d \v x=\bfL_g=-\frac{g\hat{\v z}}{2\pi c}\int_{S} \v E_{\rm in} \cdot d \v S,
\end{equation}
where  $C_1>\partial S, C_2>\partial S, ..., C_k>\partial S$ are homotopically equivalent charge paths. We cannot know which of the circulations of the potential $\v C_{\rm out}$ displayed in (28) is connected with ${\bm L}_g$. The potential  $\v C_{\rm out}$ is ambiguous due to its gauge-dependence and its circulation $\oint_{C}\v C_{\rm out}\cdot d \v x$ is ambiguous due to its spatial delocalisation (indistinguishability of the curve $C$). Therefore the $\bfC$-explanation is unsatisfactory. But if we consider the last equality in (28) then we conclude that ${\bm L}_g$ outside the electric solenoid is unambiguously connected with the flux of the electric field confined inside the solenoid. Since the magnetic charge and the electric flux lie in different spatial regions then they nonlocally interact to produce ${\bm L}_g$. Thus, the ${\bm E}$-explanation holds.

As a theoretical application of the  electromagnetic angular  momentum $\bfL_{g}$, consider that the magnetic charge $g$ is the elementary magnetic monopole $g_0$, i.e.,  $g=g_0$. Furthermore, consider the Dirac quantisation condition \cite{15,16}: $qg=N\hbar c/2$ (with $N$ an integer) for the case in which  $q=e$ with $e$ being the electron charge and $g=g_0$ implies the relation $g_0=e/(2\alpha)$, where $\alpha$ is the fine structure constant. Under these considerations and recalling that $\Phi_e = \pi R^2 E$ is the electric flux with $E$ being the magnitude of the uniform electric field inside the electric solenoid,  (24) yields the following relation $L_{g_0}= -e R^2E/(4\alpha c)$ for the $z$-component of the corresponding electromagnetic angular momentum.

\section {Dyon-solenoid configuration}
The program followed in  Ref. \cite{1} to study the electromagnetic angular  momentum $\bfL_q$ of the charge-solenoid configuration and the similar program developed here to study the electromagnetic angular  momentum $\bfL_g$ of the monopole-solenoid configuration can naturally be combined to study the electromagnetic angular  momentum $\bfL_{qg}$ of the dyon-solenoid configuration. The interesting point of $\bfL_{qg}$ is that besides its topological and nonlocal features, it is duality invariant.

The dyon-solenoid is formed by a dyon, a particle of mass $m$ which posses both an electric charge $q$ and a magnetic charge $g$,  continuously moving around an infinitely-long solenoid of radius $R$ which encloses both a uniform electric flux $\Phi_e$ and a uniform magnetic flux $\Phi_m$ (see Fig.~3). We will call such a solenoid the dual solenoid which is centred along the $z$-axis. The electric and magnetic fields of the dual solenoid are produced by the electric and magnetic surface current densities
\begin{equation}
\v J_{e}= \frac{c\Phi_{m}\delta(\rho-R)}{4\pi^2 R^2}\,\hat{\!\bfphi},\quad\v J_{m}=-\frac{c\Phi_{e}\delta(\rho-R)}{4\pi^2 R^2}\,\hat{\!\bfphi},
\end{equation}
where $\Phi_m = \pi R^2 B$ and $\Phi_e = \pi R^2 E$ are the magnetic and electric fluxes with $B$ and $E$ being the magnitude of the uniform electric and magnetic fields inside the dual solenoid. The currents $\v J_{m}$ and $\v J_{e}$ are steady currents: $\nabla\cdot \v J_m=0$ and $\nabla\cdot \v J_e=0$ and their corresponding electric and magnetic fields satisfy the equations
\begin{eqnarray}
\nabla\cdot \v B\!\!\!\!\!&=\!\!\!\!\!&0,\quad \nabla\times \v B=\frac{\Phi_{m}\delta(\rho-R)}{\pi R^2}\,\hat{\!\bfphi},\\
\nabla\cdot \v E\!\!\!\!\!&=\!\!\!\!\!&0,\quad \nabla\times \v E=\frac{\Phi_{e}\delta(\rho-R)}{\pi R^2}\,\hat{\!\bfphi}.
\end{eqnarray}
The first equations in (30) and (31) indicate that there are no isolated electric and magnetic charges inside the dual solenoid. The solution of (30) and (31) is given by magnetic and electric fields
\begin{equation}
\v B=\frac{\Phi_{m}\Theta(R-\rho)}{\pi R^2}\hat{\v z}, \quad \v E=\frac{\Phi_{e}\Theta(R-\rho)}{\pi R^2}\hat{\v z}.
\end{equation}
which are  confined in the dual solenoid. It follows that $\v E_{\rm out}=0$ and $\v B_{\rm out}=0$ are  the electric and magnetic fields  outside $(\rho>R)$ the dual solenoid and that  $\v E_{\rm in}=\Phi_e\hat{\v z}/(\pi R^2)$ and  $\v B_{\rm in}=\Phi_m\hat{\v z}/(\pi R^2)$ are the electric and magnetic fields inside $(\rho<R)$ the dual solenoid.

From the first equation in (30) we infer the relation $\v B = \nabla \times \v A$ where $\v A$ is the magnetic vector potential. Analogously, from the first equation in (31) we infer the relation $\v E = -\nabla \times \v C$ where $\v C$ is the electric vector potential. Using these relations in the respective second equations in (30) and (31), and adopting the Coulomb gauges $\nabla \cdot \v A=0$ and $\nabla \cdot \v C=0$, we obtain the Poisson equations
\begin{equation}
\nabla^2\v A=-\frac{\Phi_{m}\delta(\rho-R)}{\pi R^2}\,\hat{\!\bfphi},\quad \nabla^2\v C= \frac{\Phi_{e}\delta(\rho-R)}{\pi R^2}\,\hat{\!\bfphi},
\end{equation}
whose solutions read
\begin{eqnarray}
\v A=\frac{\Phi_{m}}{2 \pi}\bigg[\frac{\Theta(\rho-R)}{\rho}+\frac{\rho\,\Theta(R-\rho)}{R^2}\bigg]\,\hat{\!\bfphi},\;\quad
\v C=-\frac{\Phi_{e}}{2 \pi}\bigg[\frac{\Theta(\rho-R)}{\rho}+\frac{\rho\,\Theta(R-\rho)}{R^2}\bigg]\,\hat{\!\bfphi}.
\end{eqnarray}
These potentials are not defined at $\rho=R$ because of the discontinuity of the Heaviside step function. However, a simple regularisation gives $\v A(R)= \Phi_{m}\hat{\bfphi}/(2\pi R)$ and $\v C(R)= -\Phi_{e}\hat{\bfphi}/(2\pi R),$ indicating that both $\v A$ and $\v C$ are continuous at $\rho=R.$
\begin{figure}
	\centering
	\includegraphics[scale=0.35]{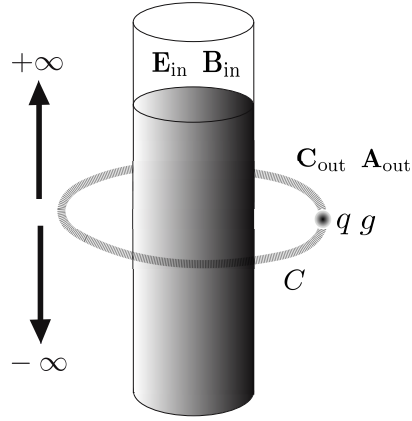}
	\caption{\normalsize{Dyon-solenoid configuration. A dyon with electric and magnetic charges is moving in the $x$-$y$ plane along the path $C$ which encircles an infinitely-long dual solenoid enclosing uniform electric and magnetic fields.}}
	\label{Fig3}
\end{figure}
We observe that the fields, potentials, currents, and fluxes associated to the dual solenoid are invariant under the electromagnetic duality transformations
\begin{eqnarray}
\nonumber\v E \to \v B,\;\, \v B\to-\v E,\,\v A\to \v C,\; \v C\to-\v A,\,\,\quad\\
\v J_{e}\to \v J_{m},\;\,\v J_{m}\to -\v J_{e},\; \Phi_{e}\to \Phi_{m},\,
\Phi_{m}\to -\Phi_{e}.
\end{eqnarray}
From (34) it follows that
\begin{equation}
\v A_{\rm out}= \frac{\Phi_m}{2 \pi\rho}\,\hat{\!\bfphi},\quad \v C_{\rm out}= -\frac{\Phi_e}{2 \pi\rho}\,\hat{\!\bfphi},
\end{equation}
where $\v A_{\rm out}$ and $\v C_{\rm out}$ are the magnetic and electric vector potentials outside the dual solenoid, which are  connected with their respective magnetic and electric fields:  $\v B_{\rm out}=\nabla \times \v A_{\rm out}=0$ and $\v E_{\rm out}=-\nabla \times \v C_{\rm out}=0$. Analogously, from (34) it follows
\begin{equation}
\v A_{\rm in}=  \frac{\rho\Phi_{m}}{2 \pi R^2}\hat{\bfphi}, \quad \v C_{\rm in}=-\frac{\rho\Phi_{e}}{2 \pi R^2}\hat{\bfphi}.
\end{equation}
where $\v A_{\rm in}$ and $\v C_{\rm in}$ are the magnetic and electric vector potentials inside the dual solenoid, which are  connected with their respective magnetic and electric fields by $\v B_{\rm in}=\nabla\times \v A_{\rm in}= \Phi_{m}\hat{\v z}/(\pi R^2)\hat{\v z}$ and $\v E_{\rm in}=-\nabla\times \v C_{\rm in}= \Phi_{e}\hat{\v z}/(\pi R^2).$ Notice that $\v A_{\rm out}$ and $\v C_{\rm out}$ are pure gauge potentials: $\v A_{\rm out}= \nabla \chi_m$ and $\v C_{\rm out}= \nabla \chi_e,$ where $\chi_m = \Phi_{m}\phi/(2 \pi)$ and $\chi_e= -\Phi_{e}\phi/(2 \pi)$ are multi-valued functions of the azimuthal coordinate: $\chi_m\neq \chi_m(\phi+2\pi)$ and $\chi_e\neq \chi_e(\phi+2\pi)$ and  satisfy $\nabla^2\chi_m=0$ and $\nabla^2\chi_e=0.$ Their gradients $\nabla\chi_m$ and $\nabla\chi_e$ are singular functions.

The region outside the dual solenoid is a region free of electromagnetic forces. The generalised Lorentz force is given by \cite{7}: $\mathbf{F} = q \dot{\v x}\times \v B/c-g \dot{\v x}\times \v E/c$, where $\dot{\v x}=d\v x/dt$ is the velocity of the dyon and $\v x$ is its position. Using (32) and $\dot{\v x} = \dot{\rho}\,\hat{\!\bfrho} + \rho \dot{\phi}\,\hat{\!\bfphi}+ \dot{z}\hat{\v z}$ in the generalised Lorentz force, we obtain
\begin{equation}
\v F  = \frac{(q\Phi_m- g\Phi_e) \Theta(R-\rho)}{\pi R^2}(\rho \dot{\phi}\,\hat{\!\bfrho}-\dot{\rho}\,\hat{\!\bfphi}).
\end{equation}
In the dyon-solenoid configuration the moving dyon is outside the solenoid $(\rho>R)$ and therefore $\Theta=0$, which implies $\v F = 0.$ In short: the generalised Lorentz force vanishes because the electric and magnetic fields are zero in the region where the dyon is moving.

\section{Topology and nonlocality of the circulation of $\v A_{\rm out} + \v C_{\rm out}$}
Let us now apply (11) to the dual solenoid by first making the identification $K=\Phi_m-\Phi_e$ in $K\,\hat{\!\bfphi}/(2\pi\rho)$ to obtain the relation
\begin{eqnarray}
\frac{K\,\hat{\!\bfphi}}{2\pi\rho}=\bigg(\frac{\Phi_m}{2 \pi \rho}-\frac{\Phi_e}{2 \pi \rho}\bigg)\,\hat{\!\bfphi}=\v A_{\rm out}+\v C_{\rm out}.
\end{eqnarray}
Since the dual solenoid encloses a singularity at $\rho=0$ then it follows from (11) that
\begin{eqnarray}
\oint_C\!(\v A_{\rm out}+\v C_{\rm out})\!\cdot \!d \v x=\left\{\begin{array}{@{}l@{\quad}l}
       n(\Phi_m-\Phi_e) & \mbox{if $C$ encloses the dual solenoid} \\[\jot]
      \,0 & \mbox{otherwise}
    \end{array}\right.
\end{eqnarray}
 The circulation of
$\v A_{\rm out}+\v C_{\rm out}$  in (40) is constant and therefore it is insensitive to the form of the curve $C$ and also to the dynamics that we can associate to this curve. Accordingly, if we consider $C_1,C_2...C_k$ different curves enclosing $n$ times  the dual solenoid then we have the equalities
\begin{eqnarray}
\oint_{C_1}(\v A_{\rm out}+\v C_{\rm out}) \cdot d \v x = \oint_{C_2}(\v A_{\rm out}+\v C_{\rm out})\cdot \!d \v x=...=
\oint_{C_k}(\v A_{\rm out}+\v C_{\rm out})\cdot d \v x.
\end{eqnarray}
The curves $C_1,C_2...C_k$ are homotopically equivalent. Therefore we do not know if $n(\Phi_m-\Phi_e$)
is connected with the circulation of $\v A_{\rm out}+\v C_{\rm out}$ along $C_1$ or along $C_2$ or along  $C_k$.
\begin{figure}
	\centering
	\includegraphics[scale=0.6]{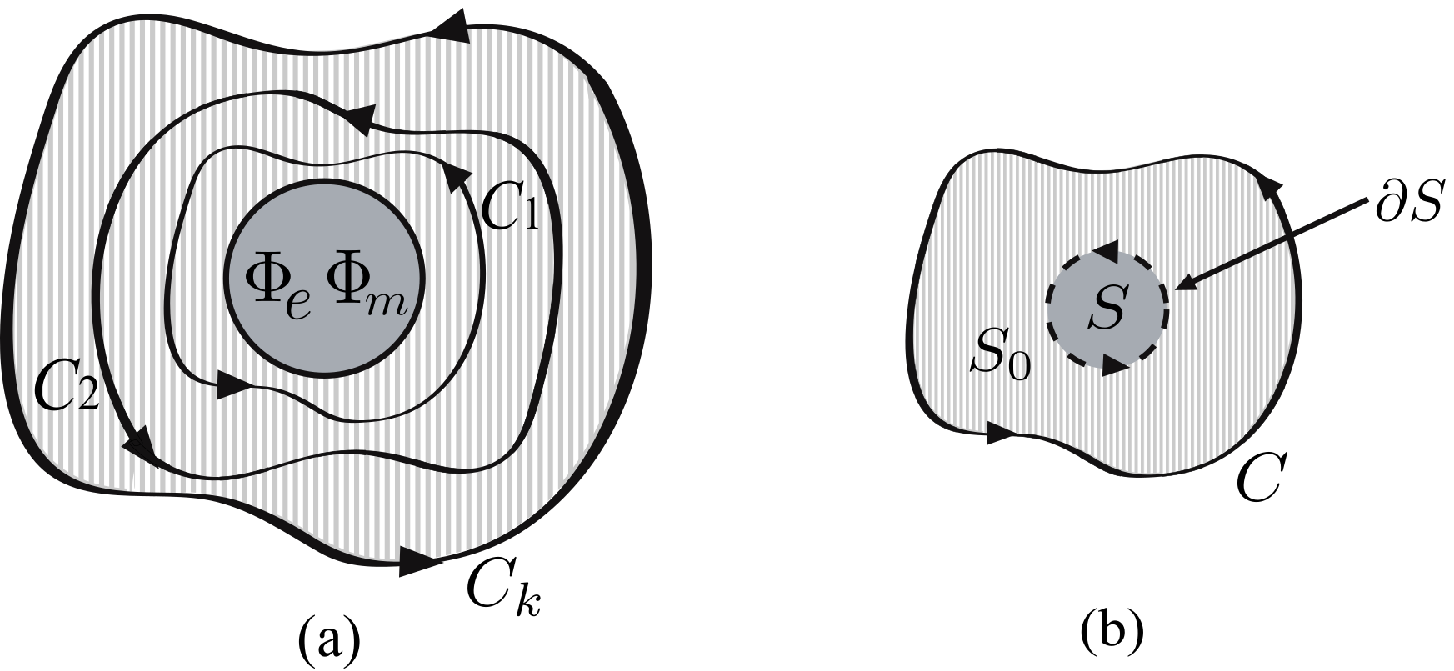}
	\caption{\normalsize{(a) The circulation of $\v A_{\rm out}+\v C_{\rm out}$ in a closed path around  the dual solenoid is insensitive to the form of the path. (b) The circulation of $\v A_{\rm out}+\v C_{\rm out}$ is taken along a path $C$ greater than the boundary $\partial S$ of the surface $S$ of the dual solenoid. From $\oint_{C>\partial S}(\v A_{\rm out}+\v C_{\rm out})\cdot d \v x=
\int_S  (\v B_{\rm in}-\v E_{\rm in}) \cdot d \v S$ it follows that the circulation of $\v A_{\rm out}+\v C_{\rm out}$  is spatially delocalised from the surface of the dual solenoid where the fluxes of the fields $\v B_{\rm in}$ and $\v E_{\rm in}$ are localised.}}
	\label{Fig4}
\end{figure}

The Stokes theorem for the charge-solenoid configuration reads $\oint_{C>\partial S}\v A_{\rm out}\cdot d \v x =\int_{S}\nabla \times \v A_{\rm in}\cdot d\v S$ and for the monopole-solenoid configuration this theorem reads $\oint_{C>\partial S}\v C_{\rm out}\cdot d \v x =\int_{S}\nabla \times \v C_{\rm in}\cdot d\v S$. It follows that the Stokes theorem applied to the dyon-solenoid configuration takes the form
\begin{equation}
\oint_{C>\partial S}(\v A_{\rm out}+\v C_{\rm out})\cdot d \v x =\int_{S}\nabla \times (\v A_{\rm in}+\v C_{\rm in})\cdot d\v S,
\end{equation}
which admits a nonlocal interpretation: while the left-hand side of (42) is defined outside the dual solenoid, its right-hand side is defined inside this solenoid.
Since $\nabla \times \v A_{\rm in}=\v B_{\rm in},$ and $\nabla \times \v C_{\rm in}=-\v E_{\rm in},$ then (42) can also be written as
\begin{equation}
\oint_{C>\partial S}(\v A_{\rm out}+\v C_{\rm out})\cdot d \v x = \int_{S}(\v B_{\rm in}-\v E_{\rm in})\cdot d\v S,
\end{equation}
according to which the difference of the magnetic and electric fluxes confined in the dual solenoid and the circulation of the sum of the magnetic and electric vector potentials along a curve outside the dual solenoid are nonlocally connected. Equations (41) and (43) imply
\begin{equation}
\oint_{C_k>\partial S}\!(\v A_{\rm out}+\v C_{\rm out})\cdot d \v x=...=\oint_{C_2>\partial S} \!(\v A_{\rm out}+\v C_{\rm out})\cdot d \v x=\oint_{C_1>\partial S} \!(\v A_{\rm out}+\v C_{\rm out})\cdot d \v x\!=\!\int_{S}(\v B_{\rm in}-\v E_{\rm in})\cdot d\v S.
\end{equation}
We have here an ambiguity because we cannot distinguish if the difference of the fluxes is connected with the circulation of the sum of the potentials
along $C_1>\partial S$ or along $C_2>\partial S$ or along  $C_k>\partial S$, i.e., the circulations in (44) are spatially delocalised with respect to the fluxes. The sum  $\v A_{\rm out}\!+\v C_{\rm out}$ is ambiguous due to the gauge-dependence of both potentials and its circulation $\oint_{C} (\v A_{\rm out}\!+\v C_{\rm out})\cdot d \v x$ is also ambiguous due to its spatial delocalisation (indistinguishability of the curve $C$).

\section{Electromagnetic angular momentum of the  dyon-solenoid configuration}
Considering the  Decomposition Theorem demonstrated in  Ref.~[1] and its dual version given in Sec.~4 we can formulate the following duality-invariant decomposition theorem which allows to derive the electromagnetic angular momentum of the  dyon-solenoid configuration.
\vskip 3pt
\noindent \emph{Duality-Invariant Decomposition Theorem.} Let $\bfcalE(\v x,t)$ and  $\bfcalB(\v x,t)$ time-dependent electric and magnetic fields and $\varrho_e(\v x,t)=\nabla\cdot \bfcalE/(4\pi)$ and $\varrho_m(\v x,t)=\nabla\cdot \bfcalB/(4\pi)$ their associated electric and magnetic charge densities.
Let $\v E(\v x)$ and $\v B (\v x)$ be time-independent electric and magnetic fields and $\v A (\v x)$ and $\v C(\v x)$ their associated vector potentials
$\nabla\times \v A=\v B$ and $\nabla\times \v C=-\v E$. The fields $\bfcalE$ and $\bfcalB$  are independent of the fields
$\v E$ and $\v B$, i.e., both sets of fields are produced by different sources. The Poynting formula for the electromagnetic angular momentum in the volume $V$ originated by the interaction between the fields $\bfcalE$ and $\bfcalB$  and the fields $\v E$ and $\v B$ is given by
\begin{equation}
\textbf{\bfL}_{\rm P}=\frac{1}{4\pi c}\int_{V}\v x \times (\v E\times \bfcalB + \bfcalE\times \v B)\,d^3x,
\end{equation}
and can be decomposed as
\begin{equation}
\textbf{\bfL}_{\rm P}=\textbf{\bfL}_{\rm M}+\textbf{\bfL}_{\rm R}+\textbf{\bfL}_{\rm G}+\textbf{\bfL}_{\rm S},
\end{equation}
where
\begin{eqnarray}
\textbf{\bfL}_{\rm M}\!\!\!\!\!&=&\!\!\!\!\!\frac{1}{c}\int_{V}\v x \times( \varrho_e\v A+\varrho_m \mathbf{C})\,d^3x, \\
\textbf{\bfL}_{\rm R}\!\!\!\!\!&=&\!\!\!\!\!\frac{1}{4\pi c}\int_{V}\mathbf{x} \times \big[(\nabla\times\bfcalE)\times \v A +(\nabla\times\bfcalB)\times \v C\big]\,d^3x,\\
\textbf{\bfL}_{\rm G}\!\!\!\!\!&=&\!\!\!\!\!\frac{1}{4\pi c}\int_{V}\mathbf{x} \times [\bfcalE(\nabla\cdot \v A)+\bfcalB(\nabla\cdot \v C)]\,d^3x,\\
\textbf{\bfL}_{\rm S}\!\!\!\!\!&=&\!\!\!\!\!\frac{1}{4\pi c}\oint_{S} \mathbf{x} \times\big[\hat{\textbf{n}}(\bfcalE\cdot
\textbf{A}+\bfcalB\cdot\textbf{C})-\textbf{A}(\hat{\textbf{n}}\cdot\bfcalE)-\textbf{C}(\hat{\textbf{n}}\cdot\bfcalB)-\bfcalE(\hat{\textbf{n}}\cdot \textbf{A})-\bfcalB(\hat{\textbf{n}}\cdot \textbf{C})\big]\,d{S}.
\end{eqnarray}
Here $S$ is the surface of the volume $V$. The proof of this theorem is given in two parts. In  Appendix A we show the validity of (45). In Appendix B we show that (46) follows from a tensor identity. The term $\textbf{\bfL}_{\rm R}$ in (48) deals with possible radiative effects. The term $\textbf{\bfL}_{\rm G}$ in (49) involving
$\nabla\cdot \v A$ and $\nabla\cdot \v C$ deals with the adopted gauge for the potentials $\v A$ and $\v C$. The term $\textbf{\bfL}_{\rm S}$ in (50) is a surface term.

We can apply (46) to the dyon-solenoid configuration where the volume $V$ covers all space except the volume of the dual solenoid. In this case we immediately conclude that $\textbf{\bfL}_{\rm P}=0$ because $\textbf{E}=0$ and  $\textbf{B}=0$ outside the dual solenoid. At the end of Appendix E  we show that $\textbf{\bfL}_{\rm R}=0$, which indicates that the electromagnetic angular momentum is insensitive to the radiative effects of the encircling dyon. We have  $\textbf{\bfL}_{\rm G}=0$ because $\nabla\cdot \v A_{\rm out}=0$ and
 $\nabla\cdot \v C_{\rm out}=0$. Thus (46) reduces to $0=\textbf{\bfL}_{\rm M}+\textbf{\bfL}_{\rm S}$, indicating that there exists $\textbf{\bfL}_{\rm M}$ in the volumetric space  where the dyon is moving around the dual solenoid and there exists $\textbf{\bfL}_{\rm S}$ on the surface of this volumetric space which lies at infinity. In the relation $0=\textbf{\bfL}_{\rm M}+\textbf{\bfL}_{\rm S}$  the pieces $\textbf{\bfL}_{\rm M}$ and $\textbf{\bfL}_{\rm S}$ are defined in different spatial regions.

Equation (47) allows us to obtain the duality-invariant  electromagnetic angular momentum $\bfL_{qg}$ of the dyon-solenoid configuration covering all space except the surface $S$ which lies at infinity. We assume that the dyon is localised in the $x$-$y$ plane and therefore its position vector is given by
\begin{equation}
\v x_{qg}(t)=\{\rho_{qg}(t)\cos\phi_{qg}(t),\rho_{qg}(t)\sin\phi_{qg}(t),0\}=\rho_{qg}(t)\,\hat{\!\bfrho}.
\end{equation}
The  associated electric and magnetic charge densities are
 \begin{eqnarray}
 \varrho_m(\v x',t)=\frac{q}{\rho'} \delta\{\rho' - \rho_{qg}(t)\}\delta\{\phi' - \phi_{qg}(t)\}\delta\{z'\}, \\
\varrho_e(\v x',t)=\frac{g}{\rho'} \delta\{\rho' - \rho_{qg}(t)\}\delta\{\phi' - \phi_{qg}(t)\}\delta\{z'\}.
\end{eqnarray}
 The vector potentials are $\v A_{\rm out}(\v x')=\Phi_m\, \hat{\!\bfphi}/(2\pi\rho')$ and $\v C_{\rm out}(\v x')=-\Phi_e\, \hat{\!\bfphi}/(2\pi\rho').$ A generic point reads $\v x' = \rho'\hat{\bfrho} +z'\hat{\v z}$. Using these ingredients we integrate the right-hand side of (47), obtaining
\begin{equation}
\frac{1}{c}\int_{V}\v x' \times (\varrho_e(\v x',t)\v A_{\rm out}(\v x')-\varrho_m(\v x',t)\v C_{\rm out}(\v x'))\,d^3x'=\frac{n(q\Phi_m- g\Phi_e)}{2\pi c}\hat{\v z},
\end{equation}
where $n$ is the winding number representing the number of times the dyon travels its closed path around the dual solenoid. In the derivation of (54) we have used $\int^{\infty}_{-\infty}\delta(z')dz'=1,\int^{\infty}_{-\infty}z'\delta(z')dz'=0,
\int^{\infty}_{R}\delta\{\rho'-\rho_{qg}(t)\}\,d\rho'=\Theta\{\rho_{qg}(t) - R\}=1\,(\rho_{qg}>R)$ and $\oint_{C} \delta\{\phi' - \phi_{qg}(t)\}\,d\phi'=n$.
Therefore (54) represents the accumulated electromagnetic angular  momentum of the dyon-solenoid configuration
\begin{equation}
\bfL_{qg}=\frac{n(q\Phi_m- g\Phi_e)}{2\pi c}\hat{\v z},
\end{equation}
which states that every time the dyon goes around the dual solenoid, the dyon-solenoid configuration acquires the electromagnetic angular  momentum $\bfL_{qg}^{(n=1)}\!=\!(q\Phi_m- g\Phi_e)\hat{\v z}/(2\pi c)$ and therefore after $n$ times this configuration accumulates the electromagnetic angular  momentum given by (55), which can compactly be written as $\bfL_{qg}=n\bfL_{qg}^{(n=1)}$. Equations (40) and (55) yield
\begin{equation}
\bfL_{qg}=\frac{\hat{\v z}}{2\pi c}\oint_{C} (q\v A_{\rm out}+g\v C_{\rm out})\cdot d \v x,
\end{equation}
which is gauge invariant on account of the gauge invariance of the circulation of the potentials  $\v A_{\rm out}$ and $\v C_{\rm out}.$ The accumulated electromagnetic angular  momentum ${\bm L}_{qg}$ depends on the circulation of the potentials outside the dual solenoid.  Using (43) we can see that (56) can be expressed as
\begin{equation}
\bfL_{qg}= \frac{\hat{\v z}}{2\pi c}\int_{S}(q\v B_{\rm in}-g\v E_{\rm in})\cdot d\v S,
\end{equation}
which states that the accumulated electromagnetic angular  momentum ${\bm L}_{qg}$ outside the dual solenoid depends on the fluxes of the electric and magnetic fields inside this dual solenoid.

As a theoretical application of the  electromagnetic angular  momentum $\bfL_{qg}$, consider a
dyon having the elementary charges $q=q_0$ and $g=g_0$. Furthermore, consider the Schwinger-Zwanziger
quantisation condition \cite{7,18}: $q_1g_2 - q_2g_1=N\hbar c/2$ ($N$ integer), with a solution given by $q = n_e q_0$
and $g=n_gg_0,$ where $n_e$ and $n_g$ are integers, $q_0=e$ where $e$ is the electron's charge and
$g_0 = e/(2 \alpha)$ where $\alpha$ is the fine structure constant. Therefore using the elementary
charges $q_0=e$ and $g_0=e/(2\alpha)$ together with the fluxes $\Phi_e = \pi R^2 E$  and
$\Phi_m=\pi R^2 B$ inside the dual solenoid we obtain $L_{q_0 g_0}=eR^2[B-E/(2\alpha)]/(2c)$ for
the $z$-component of the corresponding electromagnetic angular  momentum.

\section{Potential interpretation vs field interpretation}
Equations (56) and (57) suggest two different interpretations of the physical origin of the electromagnetic angular  momentum $\bfL_{qg}$. Let us combine both equations to obtain the relation
\begin{equation}
\frac{\hat{\v z}}{2\pi c}\oint_{C}(q\v A_{\rm out}+g\v C_{\rm out}) \cdot d \v x=\bfL_{qg}=\frac{\hat{\v z}}{2\pi c}\int_{S}(q\v B_{\rm in}-g\v E_{\rm in})\cdot d\v S.
\end{equation}
The first interpretation, which we will call the potential-explanation, is supported by the first equality in (58). This states that $\v A_{\rm out}$ and $\v C_{\rm out}$ locally act through their circulations on the dyon originating $\bfL_{qg}$. In other words: $\v A_{\rm out}$ and $\v C_{\rm out}$ exist in each point of the trajectory of the dyon and therefore they locally act on the dyon producing $\bfL_{qg}$.
The second interpretation, which we will call the field-explanation, is supported by the second equality in (58), which states that the fields
 $\v B_{\rm in}$ and  $\v E_{\rm in}$ nonlocally act through their fluxes on the dyon originating $\bfL_{qg}$ because the dyon and the fields $\v B_{\rm in}$ and $\v E_{\rm in}$ lie in different regions. Put in other words: $\v B_{\rm in}$ and $\v E_{\rm in}$ exist inside the dual solenoid and not in each point of the trajectory of the dyon outside the solenoid and therefore $\v B_{\rm in}$ and $\v E_{\rm in}$ nonlocally act on the dyon producing $\bfL_{qg}$. We do not support the potential-explanation for the following arguments: the potentials $\v A_{\rm out}$ and $\v C_{\rm out}$ are gauge-dependent and therefore the quantity $q\v A_{\rm out}+g\v C_{\rm out}$ cannot represent a physical quantity. The Stokes theorem applied to the dyon-solenoid configuration takes the form (see (44))
\begin{eqnarray}
\frac{\hat{\v z}}{2\pi c}\oint_{C_k>\partial S}\!(q\v A_{\rm out}+g\v C_{\rm out})\cdot d \v x=...=\frac{\hat{\v z}}{2\pi c}\oint_{C_2>\partial S} \!(q\v A_{\rm out}+g\v C_{\rm out})\cdot d \v x\nonumber\\
=\frac{\hat{\v z}}{2\pi c}\oint_{C_1>\partial S} \!(q\v A_{\rm out}+g\v C_{\rm out})\cdot d \v x=\bfL_{qg}
=\frac{\hat{\v z}}{2\pi c}\int_{S}(q\v B_{\rm in}-g\v E_{\rm in})\cdot d\v S,
\end{eqnarray}
where the different dyon paths $C_1>\partial S, C_2>\partial S, ..., C_k>\partial S$ are homotopically equivalent. Considering the equalities of the left-hand side of  $\bfL_{qg}$ in (59) there is a manifest ambiguity because we cannot distinguish if $\bfL_{qg}$ is connected with the circulation of $q\v A_{\rm out}+g\v C_{\rm out}$ along $C_1>\partial S$ or along $C_2>\partial S$ or along  $C_k>\partial S$. Stated differently, $q\v A_{\rm out}+g\v C_{\rm out}$  is ambiguous due to the gauge-dependence of the potentials and the circulations of $q\v A_{\rm out}+g\v C_{\rm out}$
is ambiguous due to the spatial delocalisation. Thus, the potential-explanation is unsatisfactory. We admit the field-explanation because the last equality in (59) states  that ${\bm L}_{qg}$ outside the dual solenoid is unambiguously connected with the fluxes of $\v B_{\rm in}$ and $\v E_{\rm in}$ confined inside the dual solenoid. Since the dyon as well as the magnetic and electric fields lie in different spatial regions then they nonlocally interact to produce ${\bm L}_{qg}$.
In other words, (59) tell us that the magnetic and electric fluxes do not locally affect the trajectory of the dyon.

\section{Duality invariance of the electromagnetic angular  momentum $\bfL_{\small{\bfq\bfg}}$}
The electromagnetic angular  momentum $\bfL_{qg}$ in (55) is clearly invariant under the simultaneous application of the set formed by the duality transformations of the charges: $q\to g$ and $g\to -q$ and the set formed by the duality transformations of the fluxes: $\Phi_m\to -\Phi_e$ and $\Phi_e\to \Phi_m$. Both sets of transformations are independent because the charges $q$ and $g$ of the dyon are specified independently from the fluxes $\Phi_e$ and $\Phi_m$ of the dual solenoid. However, both  sets can be  unified by means of a generalised set of $U(1)$ electromagnetic duality transformations given by
\begin{equation}
q+ig={\rm e}^{-i\theta}\big(q'+ig'\big),\quad\Phi_{e}+i\Phi_{m}={\rm e}^{-i\theta}\big(\Phi'_{e}+i\Phi'_{m}\big),
\end{equation}
where $\theta$ is an arbitrary real angle. The transformations in (60) can explicitly be written as
\begin{eqnarray}
q\!\!\!\!\!&=&\!\!\!\!\!q'\cos\theta+g'\sin\theta,\;\quad\; \,\, \,\,\quad g=-q'\sin\theta+g'\cos\theta,\\
\Phi_{e}\!\!\!\!\!&=&\!\!\!\!\!\Phi_e'\cos\theta+\Phi_{m}'\sin\theta,\;\;\quad \Phi_{m}=-\Phi_{e}'\sin\theta+\Phi_{m}'\cos\theta,
\end{eqnarray}
and their corresponding inverse transformations read
\begin{eqnarray}
q'\!\!\!\!\!&=&\!\!\!\!\!q\cos\theta-g\sin\theta,\;\quad\;\;\,\,\,\, \,\,\quad g'=q\sin\theta+g\cos\theta,\\
\Phi'_{e}\!\!\!\!\!&=\!\!\!\!\!&\Phi_{e}\cos\theta-\Phi_{m}\sin\theta,\; \,\, \,\quad\Phi'_{m}=\Phi_{e}\sin\theta+\Phi_{m}\cos\theta.
\end{eqnarray}
From (61) and (62) it follows that the quantity $q\Phi_{m}-g\Phi_{e}$ is duality invariant
\begin{equation}
q\Phi_{m}-g\Phi_{e}=q'\Phi_{m}'-g'\Phi_{e}'.
\end{equation}
Considering (65) we can show that the electromagnetic angular  momentum of the dyon-solenoid configuration is duality invariant:
$n(q\Phi_{m}-g\Phi_{e})\hat{\v z}/(2\pi c)= \bfL_{qg} =n(q'\Phi_{m}'-g'\Phi_{e}')\hat{\v z}/(2\pi c)$. Furthermore, by exploiting the arbitrariness  of the angle $\theta$, we can demonstrate that $\bfL_{qg} =n(q'\Phi_{m}'-g'\Phi_{e}')\hat{\v z}/(2\pi c)$ describes both  $\bfL_{q} =nq\Phi_{m}\hat{\v z}/(2\pi c)$ and $\bfL_{g} =-ng\Phi_{e}\hat{\v z}/(2\pi c)$.

We have two procedures to
obtain $\bfL_{q}$ from $\bfL_{qg}$. In the first procedure we assume that all dyons have the same ratio of magnetic to electric charge: $g'/q'$= constant. Since $\theta$ is arbitrary, we can write
\begin{equation}
\frac{g'}{q'}=\tan\theta,
\end{equation}
which implies $\theta=\tan^{-1}(g'/q').$ The condition (66) and the second transformation in (61) imply the vanishing of the magnetic charge  of the dyon $g=0$ and therefore the transformations in (63) become
\begin{equation}
q'=q\cos\theta, \quad g'=q\sin\theta.
\end{equation}
(alternatively, we can demand that $g=0$ and then the transformations in (63) imply the relations in (67) which yields (66)). The relations in (67) combine with the transformations in (64) to obtain
\begin{equation}
q'\Phi'_{m}-g'\Phi'_{e}= q(-\Phi_{e}'\sin\theta+\Phi_{m}'\cos\theta)=q \Phi_{m},
\end{equation}
and making use of this relation it follows that $\bfL_{qg}$ becomes $\bfL_{q}$,
\begin{equation}
\bfL_{qg}=\Bigg[\frac{n(q'\Phi'_m-g'\Phi'_e)\hat{\v z}}{2\pi c}\Bigg]_{g'=q'\tan\theta}= \frac{nq \Phi_m}{2\pi c}\hat{\v z}=\bfL_{q}.
\end{equation}
In the second procedure we consider the ratio of the electric to magnetic fluxes which is a constant quantity. Since $\theta$ is arbitrary, we can write
\begin{equation}
\frac{\Phi'_{e}}{\Phi'_{m}}=-\tan\theta,
\end{equation}
which implies $\theta = \tan^{-1}(-\Phi'_{e}/\Phi'_{m})$. The condition (70) and the first transformation in (62) imply the vanishing of the electric flux $\Phi_{e}=0$ of the dual solenoid. Thus the transformations in (64) become
\begin{equation}
\Phi'_{e}=-\,\Phi_{m}\sin\theta, \quad \Phi'_{m}=\Phi_{m}\cos\theta,
\end{equation}
which combine the transformations in (63) to yield the result
\begin{equation}
q'\Phi'_{m}-g'\Phi'_{e}=(q'\cos\theta+g'\sin\theta)\Phi_{m}=q\Phi_{m},
\end{equation}
which implies that $\bfL_{qg}$ becomes again $\bfL_{q}$
\begin{equation}
\bfL_{qg}=\Bigg[\frac{n(q'\Phi'_{m}-g'\Phi'_{e})\hat{\v z}}{\hbar c}\Bigg]_{\Phi'_{e}=-\Phi'_{m}\tan{\theta}} = \frac{nq \Phi_m}{2\pi c}\hat{\v z}=\bfL_{q}.
\end{equation}
Analogously, we have two ways to obtain $\bfL_{g}$ from $\bfL_{qg}$. In the first way we assume again that all dyons satisfy the condition $g'/q'$= constant. Considering that the angle $\theta$ is arbitrary, we can write
\begin{equation}
\frac{g'}{q'}=-\cot{\theta},
\end{equation}
which implies $\theta = \cot^{-1}(-g'/q').$ The relation (74) and the first transformation in (61) imply  the vanishing of the electric charge of the dyon  $q=0$ and therefore the transformations in (63) become
\begin{equation}
q'=-g\sin\theta, \quad g'=g\cos\theta.
\end{equation}
Using these relations and the transformation in (64) we obtain
\begin{equation}
q'\Phi'_{m}-g'\Phi'_{e}= -g(\Phi_{e}'\cos\theta+\Phi_{m}'\sin\theta)=-g \Phi_{e},
\end{equation}
which is used in (65) to show that $\bfL_{qg}$ becomes  $\bfL_{g}$
\begin{equation}
\bfL_{qg}=\Bigg[\frac{n(q'\Phi'_m-g'\Phi'_e)\hat{\v z}}{2\pi c}\Bigg]_{g'=-q'\cot\theta}= -\frac{ng \Phi_e}{2\pi c}\hat{\v z}=\bfL_{g}.
\end{equation}
In the second way we choose the angle $\theta$ to satisfy
\begin{equation}
\frac{\Phi'_{e}}{\Phi'_{m}}=\cot\theta,
\end{equation}
which implies $\theta = \cot^{-1}(\Phi'_{e}/\Phi'_{m})$. The condition (78) and the second transformation in (62) imply $\Phi_{m}=0$ and thus the transformations in (64) become
\begin{equation}
\Phi'_{e}=\Phi_{e}\cos\theta\,,\quad\Phi'_{m}=\Phi_{e}\sin\theta.
\end{equation}
These relations together the transformations in (63) yield the relation
\begin{equation}
q'\Phi'_{m}-g'\Phi'_{e}= -(-q'\sin\theta+g'\cos\theta)\Phi_{e}=-g \Phi_{e},
\end{equation}
which is used in (65) to show that $\bfL_{qg}$ becomes again $\bfL_{g}$
\begin{equation}
\bfL_{qg}=\Bigg[\frac{n(q'\Phi'_m-g'\Phi'_e)\hat{\v z}}{2\pi c}\Bigg]_{\Phi'_{e}=\Phi'_{m}\cot{\theta}}= -\frac{ng \Phi_e}{2\pi c}\hat{\v z}=\bfL_{g}.
\end{equation}
We have then demonstrated that duality symmetry of the electromagnetic angular  momentum $\bfL_{qg}$ unifies the apparently dissimilar  electromagnetic angular  momenta $\bfL_{q}$ and $\bfL_{g}$. We then justify our claim that $\bfL_{g}$ is dual to $\bfL_{q}$.

\section{Interpretations of $\bfL_{\small{\bfq}}$ and $\bfL_{\small{\bfg}}$ in the light of duality}
In addition to the ${\bm B}$-explanation for the origin of  the electromagnetic angular  momentum $\bfL_{q}$, the duality symmetry of the electromagnetic angular  momentum $\bfL_{qg}$ allows us to give two further interpretations for the electromagnetic angular  momentum $\bfL_{q}$. To see this we observe that the relations in (67), which hold when $g=0$, imply the beautiful relation
\begin{equation}
q=\sqrt{q'^2+g'^2}.
\end{equation}
How should this relation be interpreted? Answer: a particle with the electric charge $q$ can be thought either as (i) a dyon with a non-vanishing electric charge $q$ and a vanishing magnetic charge $g=0$ or as (ii) a dyon with the electric charge $q'=\,q\cos\theta$ and the magnetic charge $g'=q\sin\theta$ (which imply (82)). Duality symmetry tells us that it is a matter of mere convention to adopt either (i) or (ii). Multiplying (82) by the magnetic flux $\Phi_{m}$ we obtain $q \Phi_{m}=\sqrt{q'^2+g'^2}\, \Phi_{m}$,
which implies
\begin{equation}
\frac{q \Phi_{m}}{2\pi c}\hat{\v z}=\frac{\sqrt{q'^2+g'^2}\, \Phi_{m}}{2\pi c}\hat{\v z}.
\end{equation}
Analogously, from (71) it follows
$\Phi_{m}=\sqrt{\Phi_{e}'^2+\Phi_{m}'^2}$ which implies $q \Phi_{m}=q\sqrt{\Phi_{e}'^2+\Phi_{m}'^2}$
and therefore
\begin{equation}
\frac{q \Phi_{m}}{2\pi c}\hat{\v z}=
\frac{q\sqrt{\Phi_{e}'^2+\Phi_{m}'^2}}{2\pi c}\hat{\v z}.
\end{equation}
The relations (83) and (84)  provide three equivalent interpretations for the origin of $\bfL_{q}$:
\vskip 3pt
\noindent (i) $\bfL_{q}= q \Phi_{m}\hat{\v z}/(2\pi c)$ is originated by the nonlocal action of the magnetic flux $\Phi_{m}$ on an electric charge $q$ (${\bm B}$-explanation)
\vskip 3pt
\noindent (ii) $\bfL_{q}= \sqrt{q'^2+g'^2}\, \Phi_{m}\hat{\v z}/(2\pi c)$ is originated by the nonlocal action of the magnetic flux $\Phi_{m}$ on a dyon possessing the electric charge $q'=\,q\cos\theta$ and the magnetic charge $g'=q\sin{\theta}$.
\vskip 3pt
\noindent (iii) $\bfL_{q}= q \sqrt{\Phi_{e}'^2+\Phi_{m}'^2}\hat{\v z}/(2\pi c)$ is originated by the nonlocal action of the fluxes
$\Phi'_{e}=-\,\Phi_{m}\sin\theta$ and $\Phi'_{m}=\Phi_{m}\cos\theta$ on the electric charge $q$ of a dyon.

Analogously, in addition to the ${\bm E}$-explanation for the origin of  the electromagnetic angular  momentum $\bfL_{g}$, the duality symmetry of $\bfL_{qg}$ provides two other interpretations for  $\bfL_{g}$. The way to arrive at the three interpretations for $\bfL_{g}$ is similar to that
leading to the interpretations (i)-(iii). After some calculation, we can obtain
\begin{eqnarray}
-\frac{g \Phi_{e}}{2\pi c}\hat{\v z}=-\frac{\sqrt{q'^2+g'^2}\, \Phi_{e}}{2\pi c}\hat{\v z},\\
-\frac{g \Phi_{e}}{2\pi c}\hat{\v z}=-\frac{g\sqrt{\Phi_{e}'^2+\Phi_{m}'^2}}{2\pi c}\hat{\v z}.
\end{eqnarray}
On the basis of these relations, we can conclude that
\vskip 3pt
\noindent (i$'$) $\bfL_{g}= -g \Phi_{e}\hat{\v z}/(2\pi c)$ is originated by the nonlocal action of the electric flux $\Phi_{e}$ on a magnetic charge $g$ (${\bm E}$-explanation).
\vskip 3pt
\noindent (ii$'$) $\bfL_{g}= -\sqrt{q'^2+g'^2}\, \Phi_{e}\hat{\v z}/(2\pi c)$ is originated by the nonlocal action of the electric flux $\Phi_{e}$ on a dyon possessing the electric charge $g'=\,g\cos\theta$ and the magnetic charge $q'=-g\sin{\theta}$.
\vskip 3pt
\noindent (iii$'$) $\bfL_{g}= -g \sqrt{\Phi_{e}'^2+\Phi_{m}'^2}\hat{\v z}/(2\pi c)$ is originated by the nonlocal action of the fluxes
$\Phi'_{m}=\,\Phi_{e}\sin\theta$ and $\Phi'_{e}=\Phi_{e}\cos\theta$ on the magnetic charge $g$ of a dyon.

\section{Does the radiation from the dyon affect the electromagnetic angular  momentum $\bfL_{\small{\bfq\bfg}}$?}
On the basis of the discussion given here for the electromagnetic angular  momentum $\bfL_{qg}$, the answer to the question in the title of this section is in the negative. In fact, we have seen that $\bfL_{qg}$ being a topological invariant does not depend on the dynamics of the encircling dyon which implies that $\bfL_{qg}$ is insensitive to the electric and magnetic fields produced by the motion of the encircling dyon and in particular to their radiation fields. This is particularly striking. The fact that the radiation emitted by the dyon does not affect $\bfL_{qg}$ is hard to understand from a physical point of view. In our formal treatment this means that the piece $\bfL_{R}$ of  $\bfL_{qg}$ given by (48) must be zero. However, the explicit demonstration that  $\bfL_{R}=0$ is extremely laborious because this radiative piece involves the Li\'enard-Wiechert fields of the encircling dyon whose determination turns out to be too cumbersome ---as is well-known, there are few nontrivial cases for which the Li\'enard-Wiechert fields can explicitly be calculated. In Appendix C we obtain the Li\'enard-Wiechert fields of a non-relativistic dyon in uniform circular motion.
As expected, these fields are shown to depend on an expression for the  retarded time which implicitly depends on the retarded time itself. However, in Appendix D we show how we can eliminate the implicit dependence of the retarded time and use this result in Appendix E to show that $\bfL_{R}=0$ when the dyon is in uniform circular motion. It is pertinent to note that Appendices C, D and E describe long and laborious calculations. They are required in our treatment, mainly for consistency reasons. The short answer to the question in the title this section is: no because $\bfL_{\small{\bfq\bfg}}$ is a topological quantity.
\section{Discussion: quantum mechanics enters into scene }
We have seen that if the curve $C$ encircles $n$ times an electric solenoid then topology and classical electrodynamics conspire to produce the relation (15): $\oint_{C>\partial S}\v C_{\rm out}\cdot d \v x = -\int_{S}\v E_{\rm in}\cdot d\v S.$ Let us multiply this relation by the non-vanishing constant ${\cal K}_1$,
\begin{equation}
{\cal K}_1\oint_{C>\partial S}\v C_{\rm out}\cdot d \v x = -{\cal K}_1\int_{S}\v E_{\rm in}\cdot d\v S.
\end{equation}
If ${\cal K}_1=g/(2\pi c)$ then (87) becomes the $z$-component of the electromagnetic angular  momentum $\bfL_g$:
\begin{equation}
L_g=\frac{g}{2\pi c}\oint_{C>\partial S}\v C_{\rm out}\cdot d \v x = -\frac{q}{2\pi c}\int_{S}\v E_{\rm in}\cdot d\v S=-\frac{ng\Phi_e}{2\pi c},
\end{equation}
If ${\cal K}_1=g/(\hbar c)$ then (87) becomes the dual of the Aharonov-Bohm (DAB) phase \cite{5,6}:
\begin{equation}
\delta_{\rm DAB}=\frac{g}{\hbar c}\oint_{C>\partial S}\v C_{\rm out}\cdot d \v x = -\frac{g}{\hbar c}\int_{S}\v E_{\rm in}\cdot d\v S=-\frac{ng\Phi_e}{\hbar c}.
\end{equation}
Accordingly,  topology, classical electrodynamics and quantum mechanics conspire to produce the DAB phase. Both the classical quantity $L_g$ and the quantum quantity $\delta_{\rm DAB}$  satisfy the linear relation
\begin{equation}
\delta_{\rm DAB}=\frac{2\pi}{\hbar}L_g,
\end{equation}
which exhibits the same form than that corresponding to the charge-solenoid configuration \cite{3,4}: $\delta_{\rm AB}=2\pi L_q/\hbar$.
Both $L_g$ and $\delta_{\rm DAB}$ are determined by the circulation of $\textbf{C}_{\rm out}$ whose topological and nonlocal features are implicit in (87). It follows that these features are translated to both the classical quantity $L_g$ and the quantum quantity $\delta_{\rm DAB}$ and therefore the claim that the former should be considered the classical counterpart of the latter naturally arises and (90) supports this claim.

By the same token, if the curve $C$ encircles $n$ times a dual solenoid then topology and electrodynamics with magnetic charges work together to produce the duality-invariant relation (43): $\oint_{C>\partial S}(\v C_{\rm out}+\v C_{\rm out})\cdot d \v x = \int_{S}(\v B_{\rm in}-\v E_{\rm in})\cdot d\v S,$ which is then multiplied by the non-vanishing constant ${\cal K}_2$ to obtain the relation
\begin{equation}
{\cal K}_2\oint_{C>\partial S}(q\v A_{\rm out}+g\v C_{\rm out}) \cdot d \v x=\bfL_{qg}={\cal K}_2\int_{S}(q\v B_{\rm in}-g\v E_{\rm in})\cdot d\v S.
\end{equation}
If ${\cal K}_2=1/(2\pi c)$ then (91) becomes the $z$-component of the electromagnetic angular  momentum $\bfL_{qg}$:
\begin{equation}
L_{qg}=\frac{1}{2\pi c}\oint_{C>\partial S}(q\v A_{\rm out}+g\v C_{\rm out})\cdot d \v x = \frac{1}{2\pi c}\int_{S}(q\v B_{\rm in}-g\v E_{\rm in})\cdot d\v S=     \frac{n(q\Phi_m-g\Phi_e)}{2\pi c},
\end{equation}
If ${\cal K}_2=1/(\hbar c)$ then (91) becomes the duality-invariant quantum phase \cite{8}:
\begin{equation}
\delta_{\rm D}=\frac{1}{\hbar c}\oint_{C>\partial S}(q\v A_{\rm out}+g\v C_{\rm out})\cdot d \v x = \frac{1}{\hbar c}\int_{S}(q\v B_{\rm in}-g\v E_{\rm in})\cdot d\v S=\frac{n(q\Phi_m-g\Phi_e)}{\hbar c}.
\end{equation}
Therefore, topology, classical electrodynamics and quantum mechanics conspire to produce the duality-invariant quantum phase $\delta_{\rm D}$.
Both the classical quantity $L_{qg}$ and the quantum quantity $\delta_{\rm D}$  satisfy the linear relation
\begin{equation}
\delta_{\rm D}=\frac{2\pi}{\hbar}L_{qg}.
\end{equation}
Since both $L_{qg}$ and $\delta_{\rm D}$ are given in terms of the circulations of $\textbf{A}_{\rm out}$ and
$\textbf{C}_{\rm out}$ and these involve the topological and nonlocal features implicit in (91) then these features are translated to both the classical electromagnetic angular  momentum $L_{qg}$ and the quantum phase $\delta_{\rm D}$. Therefore we claim that former should be considered the classical counterpart of the latter which is supported by (94).

\section{Conclusion}
 In this paper we have extended some of the ideas presented in \cite{1} concerning the topology and nonlocality of the electromagnetic angular  momentum $\bfL_q=nq\Phi_m/(2\pi c)$. In the first part, we have proved that the dual of $\bfL_q$ reads  $\bfL_g=[g\hat{\v z}/(2\pi c)]\oint_{C>\partial S}\textbf{C}_{\rm out}\cdot d\v x=-[g\hat{\v z}/(2\pi c)]\int_{S}\v B_{\rm in}\cdot d\v S=-ng\Phi_e/(2\pi c)$. This is the
 electromagnetic angular  momentum of the configuration formed by a magnetic charge $g$ encircling an infinitely-long electric solenoid enclosing a uniform electric flux $\Phi_e$. We have shown that $\bfL_g$ is topological because it depends on a winding number and is nonlocal because the electric field inside the electric solenoid acts on the magnetic charge in regions where this field does not exist. We have argued that $\bfL_g$ should be considered the classical counterpart of the dual of the AB phase introduced in \cite{5}.  In the second part, we have showed that the electromagnetic angular  momentum
$\bfL_{qg}=[\hat{\v z}/(2\pi c)]\oint_{C>\partial S} (q\v A_{\rm out}+g\v C_{\rm out})\cdot d \v x= [\hat{\v z}/(2\pi c)]\int_{S}(\v B_{\rm in}-\v E_{\rm in})\cdot d\v S=n(q\Phi_{m}-g\Phi_{e})\hat{\v z}/(2\pi c)$ of the configuration formed by a dyon encircling an infinitely-long dual solenoid enclosing uniform electric and magnetic fluxes is topological because it depends on a winding number, is nonlocal because the electric and magnetic fields of this dual solenoid act on the dyon in regions  where these fields do not exist, and is invariant under electromagnetic duality transformations. We have shown that the duality symmetry of $\bfL_{qg}$ allows us to unify  both $\bfL_q$ and  $\bfL_g$. This symmetry also allows us to give different physical interpretations for these electromagnetic angular momenta. We have argued that $\bfL_{qg}$ should be considered as the classical counterpart of the duality-invariant quantum phase introduced in \cite{8}. We have also demonstrated, by means of an explicit calculation, that the radiative effects of the Li\'enard-Wiechert fields of the non-relativistic dyon moving in  uniform circular motion around the dual solenoid do not affect $\bfL_{qg}$. Our detailed discussion on the electromagnetic angular momenta $\bfL_g$ and $\bfL_{qg}$ has the purpose of calling attention that topology, nonlocality and duality can consistently coexist in classical electrodynamics.
\vskip 10pt
\noindent \textbf{Data Availability Statement.} No datasets were generated or analysed in this paper.

\renewcommand\theequation{A\arabic{equation}}
\setcounter {equation}{0}
\section*{Appendix A. Poynting theorem for two sets of Maxwell's equations and the proof of (45)}
Consider two independent sets of electromagnetic equations. The first set describes the time-dependent electric and magnetic fields ${\bfcalE}(\v x,t)$ and ${\bfcalB}(\v x,t)$ produced by the electric and magnetic charge densities  $\varrho_e(\v x,t)$ and $\varrho_m(\v x,t)$ and the electric and magnetic current densities $\bfcalJ_e(\v x,t)$ and $\bfcalJ_m(\v x,t)$. The corresponding Maxwell equations read
\begin{eqnarray}
\nabla\cdot{\bfcalE}=4\pi\varrho_e,\;\;
\nabla\cdot{\bfcalB}=4\pi\varrho_m,\;\;\nabla\times{\bfcalE}+\frac{1}{c} \frac{\partial{\bfcalB}}{\partial t} = -\frac{4\pi}{c}\bfcalJ_m,\;\;\nabla\times{\bfcalB}-\frac{1}{c} \frac{\partial{\bfcalE}}{\partial t} =\frac{4\pi}{c}\bfcalJ_e.
\end{eqnarray}

The second set describes the electric and magnetic fields $\v E(\v x,t)$ and $\v B(\v x,t)$ produced by the charge and current densities $\rho_e(\v x,t), \rho_m(\v x,t), \v J_e(\v x,t)$ and $\v J_m(\v x,t)$ and satisfying the Maxwell equations
\begin{eqnarray}
\nabla\cdot\v E=4\pi\rho_e,\;\;
\nabla\cdot\v B=4\pi\rho_m,\;\; \nabla\times\v E+\frac{1}{c} \frac{\partial\v B}{\partial t} =-\frac{4\pi}{c}\v J_m,\;\; \nabla\times \v B-\frac{1}{c} \frac{\partial \v E}{\partial t} =\frac{4\pi}{c}\v J_e,
\end{eqnarray}
We shall now obtain the Poynting theorem of the system formed by (A1) and (A2). Using these equations we can directly show that
the divergence to the vector $\v E\times \bfcalB + \bfcalE\times \v B$ yields the identity
\begin{equation}
\nabla\cdot (\v E\times \bfcalB + \bfcalE\times \v B)= -\frac{4\pi}{c}(\v E\cdot \bfcalJ_e  +\bfcalE\cdot \v J_e+\v B\cdot \bfcalJ_m  +\bfcalB\cdot \v J_m)-\frac{1}{c} \frac{\partial}{\partial t}\bigg( \bfcalE\cdot \v E + \bfcalB\cdot \v B \bigg),
\end{equation}
which implies the Poynting theorem
\begin{equation}
\nabla \cdot \v S + \frac{\partial U}{\partial t}=-\v E\cdot \bfcalJ_e  -\bfcalE\cdot \v J_e-\v B\cdot \bfcalJ_m  -\bfcalB\cdot \v J_m.
\end{equation}
where the interaction energy density is given by
\begin{equation}
U= \frac{1}{4\pi}\bigg( \bfcalE\cdot \v E + \bfcalB\cdot \v B\bigg),
\end{equation}
and the interaction Poynting vector by
\begin{equation}
\v S= \frac{c}{4\pi}(\v E\times \bfcalB + \bfcalE\times \v B).
\end{equation}
Equations (A4)-(A6) are of general character. In particular they hold when the electric and magnetic are time-independent $ {\v E}(\v x)$ and ${\v B}(\v x)$ and produced by the current densities ${\v J_e}(\v x)$ and ${\v J_m}(\v x)$ satisfying the time-independent equations $\nabla\cdot\v E=0,\nabla\times\v E=-4\pi\v J_m/c,\nabla\cdot\v B=0,\nabla\times \v B=4\pi\v J_e/c.$ For this case the corresponding electromagnetic momentum density is given by $ \v g=(\v E\times \bfcalB + \bfcalE\times \v B)/(4\pi c)$
and its associated electromagnetic angular momentum density by $\bfl=\v x\times(\v E\times \bfcalB + \bfcalE\times \v B)/(4\pi c),$ whose volume integration yields the electromagnetic angular momentum given in (45).

\renewcommand\theequation{B\arabic{equation}}
\setcounter {equation}{0}
\section*{Appendix B. Proof of (46)}
A direct vector calculation leads to the identities
\begin{eqnarray}
\nonumber \partial^k\big[\varepsilon^{sqi}x_q(\delta_{ik}{\cal E}_m A^m-{\cal E}_k A_i-{\cal E}_i A_k)\big]\!\!\!\!\!&=&\!\!\!\!\!-4\pi\varrho_e\varepsilon^{sqi}x_q A_i+\varepsilon^{sqi}x_q {\cal E}^k(\partial_k A_i-\partial_i A_k)\nonumber\\&&+\varepsilon^{sqi}x_q A^k(\partial_i{\cal E}_k-\partial_k{\cal E}_i)-\varepsilon^{sqi}x_q{\cal E}_i\partial^kA_k\nonumber \\
\!\!\!\!\!&=&\!\!\!\!\!-4\pi\big[\varrho_e\v x\times \v A \big]^s +\big[\v x\times(\bfcalE\times\v B) \big]^s\nonumber \\
&&-\big[\v x\times\{(\nabla\times\bfcalE)\times\v A\} \big]^s-\big[\v x\times\bfcalE(\nabla\cdot\v A)\big]^s.
\end{eqnarray}
\begin{eqnarray}
\nonumber \partial^k\big[\varepsilon^{sqi}x_q(\delta_{ik}{\cal B}_m C^m-{\cal B}_k C_i-{\cal B}_i C_k)\big]\!\!\!\!\!&=&\!\!\!\!\!-4\pi\varrho_m\,\varepsilon^{sqi}x_q C_i+\varepsilon^{sqi}x_q {\cal B}^k(\partial_k C_i-\partial_i C_k)\nonumber\\&&+\varepsilon^{sqi}x_q C^k(\partial_i{\cal B}_k-\partial_k{\cal B}_i)-\varepsilon^{sqi}x_q{\cal B}_i\partial^kC_k\nonumber \\
\!\!\!\!\!&=&\!\!\!\!\!-4\pi\big[\varrho_m\,\v x\times \v C \big]^s +\big[\v x\times(\v E\times\bfcalB) \big]^s\nonumber \\
&&-\big[\v x\times\{(\nabla\times\bfcalB)\times\v C\} \big]^s-\big[\v x\times\bfcalB(\nabla\cdot\v C)\big]^s.
\end{eqnarray}
The identities (B1) and (B2) imply
\begin{eqnarray}
\nonumber
[\v x\times(\bfcalE\times\v B +\v E\times\bfcalB)]^s\!\!\!\!\!&=&\!\!\!\!\![\v x \times( \varrho_e\v A+\varrho_m{\v C})]^s+  \big[\v x\times\{(\nabla\times\bfcalE)\times\v A+(\nabla\times\bfcalB)\times\v C\}  \big]^s\\
 &&+\;[\v x \times \{\bfcalE(\nabla\cdot \v A)+\bfcalB(\nabla\cdot \v C)\}]^s\nonumber\\
 &&+\;\partial^k\big[\varepsilon^{sqi}x_q(\delta_{ik}({\cal E}_m A^m+{\cal B}_m C^m) -{\cal E}_k A_i-{\cal B}_k C_i-{\cal E}_i A_k-{\cal B}_i C_k)\big]^s.\nonumber\\
\end{eqnarray}
Volume integration of (B3) gives
\begin{eqnarray}
\frac{1}{4\pi c}\int_{V}\,[\v x\!\!\!\!\!\!\!\!  &\times&\!\!\!\!\!\!\!\! (\bfcalE\times\v B +\v E\times\bfcalB)]^s\,d^3x\qquad\nonumber\\
 \!\!\!\!\!&=&\!\!\!\!\!\frac{1}{c}\int_{V}[\v x \times( \varrho_e\v A+\varrho_m{\v C})]^s    d^3x\nonumber \\
&&+\frac{1}{4\pi c}\int_{V}\big[\v x\times\{(\nabla\times\bfcalE)\times\v A+(\nabla\times\bfcalB)\times\v C\}  \big]^s  d^3x\nonumber\\
&&+\frac{1}{4\pi c}\int_{V}[\mathbf{x} \times [\bfcalE(\nabla\cdot \v A)+\bfcalB(\nabla\cdot \v C)]^s\,d^3x\nonumber\\
&&+\frac{1}{4\pi c}\oint_{S} [\mathbf{x} \times\{\hat{\textbf{n}}(\bfcalE\cdot
\textbf{A}+\bfcalB\cdot\textbf{C})-\textbf{A}(\hat{\textbf{n}}\cdot\bfcalE)-\textbf{C}(\hat{\textbf{n}}\cdot\bfcalB)-\bfcalE(\hat{\textbf{n}}\cdot \textbf{A})-\bfcalB(\hat{\textbf{n}}\cdot \textbf{C})\}\big]^s\,d{S},\nonumber\\
\end{eqnarray}
where the volume integral of the last term on the right-hand side of (B3) has been transformed into a surface integral by making the replacements $\partial^k\to (\hat{\v n})^k$ and $\int_{V}d^3x\to\oint_{S}dS$. Equation (B4) is equivalent to (46).

\renewcommand\theequation{C\arabic{equation}}
\setcounter{equation}{0}
\section*{Appendix C. The Li\'enard-Wiechert fields of a non-relativistic dyon in uniform circular motion}
The explicit form of the Li\'enard-Wiechert fields of an arbitrarily moving dyon were first derived by one of us \cite{19,20}. These fields describe the retarded solutions of the Maxwell's generalised equations (A1) for the case of a point dyon in arbitrary motion and can be expressed as
\begin{eqnarray}
\bfcalE\!=\! \Bigg[\frac{q(\widehat{\bfcalR}\!-\!\bfbeta)- g\bfbeta \times\widehat{\bfcalR}}{\gamma^2{\cal R}^2(1\!-\!\widehat{\bfcalR}\cdot \bfbeta)^3}+\frac{q\widehat{\bfcalR}\times((\widehat{\bfcalR}\!-\!\bfbeta)\times \dot{\bfbeta})}{c {\cal R}(1-\widehat{\bfcalR}\cdot \bfbeta)^3}+\frac{g(1\!-\!\widehat{\bfcalR}\cdot\bfbeta)\widehat{\bfcalR}\times\dot{\bfbeta} + g(\widehat{\bfcalR}\cdot\dot{\bfbeta})\widehat{\bfcalR}\times \bfbeta}{c {\cal R}(1-\widehat{\bfcalR}\cdot \bfbeta)^3}\Bigg]_{\rm r},\quad\\
\bfcalB\!=\! \Bigg[\frac{g(\widehat{\bfcalR}\!-\!\bfbeta)+ q\bfbeta \times\widehat{\bfcalR}}{\gamma^2{\cal R}^2(1-\widehat{\bfcalR}\cdot \bfbeta)^3}+\frac{g\widehat{\bfcalR}\times((\widehat{\bfcalR}\!-\!\bfbeta)\times \dot{\bfbeta})}{c {\cal R}(1-\widehat{\bfcalR}\cdot \bfbeta)^3}-\frac{q(1\!-\!\widehat{\bfcalR}\cdot\bfbeta)\widehat{\bfcalR}\times\dot{\bfbeta} +q(\widehat{\bfcalR}\cdot\dot{\bfbeta})\widehat{\bfcalR}\times \bfbeta}{c {\cal R}(1-\widehat{\bfcalR}\cdot \bfbeta)^3}\Bigg]_{\rm r},\quad
\end{eqnarray}
where $\bfcalR=\v x - \v x_{qg}(t_r),$ with $\v x$ being the field point and $\v x_{qg}$ the position of the dyon, ${\cal R}= |\v x - \v x_{qg}(t_r)|,$ $\widehat{{\bfcalR}}= \bfcalR/{\cal{R}}$, $\bfbeta= \dot{\v x}_{qg}(t_r)/c=(1/c)d \v x_{qg}/dt_r,$ $\dot{\bfbeta}(t_r)= d \bfbeta/dt_r,$ $\gamma=1/\sqrt{1-\bfbeta^2},$ and the square brackets $[\,\,\,\,\,]_{\rm r}$ indicate that the enclosed quantities are to be evaluated at the retarded time $t_r = t- {\cal R}(t_r)/c.$ The terms varying as $1/{\cal R}^2$ are the velocity fields and the terms varying as $1/{\cal R}$ are the acceleration fields. Let us now assume that the dyon is moving with non-relativistic velocity $\beta<<1$. In this case $\gamma\approx1,$ $(1-\widehat{\bfcalR}\cdot \bfbeta)\approx1,$ and $(\widehat{\bfcalR}-\bfbeta)\approx\widehat{\bfcalR},$ and the fields (C1) and (C2) reduce to the expressions
\begin{eqnarray}
\bfcalE= \Bigg[\frac{q\widehat{\bfcalR}- g\bfbeta \times\widehat{\bfcalR} }{{\cal R}^2}+\frac{q\widehat{\bfcalR}\times(\widehat{\bfcalR}\times \dot{\bfbeta})}{c {\cal R}}+\frac{g\widehat{\bfcalR}\times\dot{\bfbeta} + g(\widehat{\bfcalR}\cdot\dot{\bfbeta})\widehat{\bfcalR}\times \bfbeta}{c {\cal R}}\Bigg]_{\rm r},\\
\bfcalB= \Bigg[\frac{g\widehat{\bfcalR}+ q\bfbeta \times\widehat{\bfcalR}}{{\cal R}^2}+\frac{g\widehat{\bfcalR}\times(\widehat{\bfcalR}\times \dot{\bfbeta})}{c {\cal R}}-\frac{q\widehat{\bfcalR}\times\dot{\bfbeta} +q(\widehat{\bfcalR}\cdot\dot{\bfbeta})\widehat{\bfcalR}\times \bfbeta}{c {\cal R}}\Bigg]_{\rm r}.
\end{eqnarray}
The fields in (C3) and (C4) are suitable to obtain the fields due to a non-relativistic dyon in uniform circular motion. We assume that the dyon lies in the $x$-$y$ plane and moving in a circle of fixed radius $a$ around the origin. The position, velocity, and acceleration of the dyon at time $t$ are given by
\begin{eqnarray}
\v x_{qg}=a[\cos(\omega t)\hat{\v x} + \sin(\omega t)\hat{\v y}],\,\,\,\,\quad\\
\dot{\v x}_{qg}=\omega a[- \sin(\omega t)\hat{\v x} + \cos(\omega t)\hat{\v y}],\,\,\\
\ddot{\v x}_{qg}=-\omega^2 a[\cos(\omega t)\hat{\v x} + \sin(\omega t)\hat{\v y}],
\end{eqnarray}
where $\omega$ is the constant angular velocity. On the other hand, at time $t$ we also have
\begin{eqnarray}
\bfcalR=[x-a\cos(\omega t)]\hat{\v x} + [y-a\sin(\omega t)]\hat{\v y} + z\hat{\v z},\qquad\,\,\,\\
{\cal R}= \sqrt{x^2 + y^2 +z^2 +a^2- 2a[x\cos(\omega t) + y \sin(\omega t)]}.
\end{eqnarray}
We also note that the fields (C3) and (C4) are invariant under the dual changes $q\to g$ and $g\to - q,$ a property that will latter be used. The next step is to transform (C5)-(C9) in cylindrical coordinates and with respect to the field coordinates $\rho,\phi,$ and $z$. On the other hand, the  source coordinates specifying the position of the dyon at time $t$ read $a,\omega t,$ and $0$. Inserting $x=\rho\cos\phi,y=\rho\sin\phi,\hat{\v x}=\cos\phi\,\hat{\!\bfrho} - \sin\phi \,\hat{\!\bfphi},$ and $\hat{\v y}=\sin\phi\,\hat{\!\bfrho}+\cos\phi\,\hat{\!\bfphi}$ in (C5)-(C9) and after some simplifications we obtain
\begin{eqnarray}
\v x_{qg}=a[\cos(\phi - \omega t)\,\hat{\!\bfrho} -\sin(\phi-\omega t) \,\hat{\!\bfphi}],\qquad\qquad\qquad\\
\dot{\v x}_{qg}=\omega a[\sin(\phi - \omega t)\,\hat{\!\bfrho} +\cos(\phi-\omega t) \,\hat{\!\bfphi}],\qquad\qquad\,\,\,\,\,\,\,\\
\ddot{\v x}_{qg}=-\omega^2 a[\cos(\phi - \omega t)\,\hat{\!\bfrho} -\sin(\phi-\omega t) \,\hat{\!\bfphi}],\qquad\quad\,\,\,\,\,\,\\
\bfcalR=[\rho - a \cos(\phi-\omega t)]\,\hat{\!\bfrho} + a\sin(\phi-\omega t)\,\hat{\!\bfphi} + z \hat{\v z},\quad\\
{\cal R}= \sqrt{\rho^2 + z^2 +a^2-2 \rho a\cos(\phi-\omega t)}.\qquad\qquad\quad\,\,\,
\end{eqnarray}
Using (C10)-(C14) and performing the specified operations, we obtain
\begin{eqnarray}
\frac{\widehat{\bfcalR}}{{\cal R}^2}=\frac{[\rho - a \cos(\phi-\omega t)]\,\hat{\!\bfrho} + a\sin(\phi-\omega t)\,\hat{\!\bfphi} + z \hat{\v z}}{[\rho^2 + z^2 +a^2-2 \rho a\cos(\phi-\omega t)]^{3/2}},\quad\qquad\qquad\qquad\qquad\qquad\qquad\qquad\qquad\,\,\,\,\,\\
\frac{\bfbeta\times \widehat{\bfcalR}}{{\cal R}^2}= \frac{\omega a}{c}\frac{z\cos(\phi-\omega t)\,\hat{\!\bfrho}-z\sin(\phi-\omega t)\,\hat{\!\bfphi} - [\rho\cos(\phi-\omega t)-a] \hat{\v z}}{[\rho^2 + z^2 +a^2-2 \rho a\cos(\phi-\omega t)]^{3/2}},\quad\qquad\qquad\qquad\qquad\qquad\,\,\\
\nonumber\frac{\widehat{\bfcalR}\times(\widehat{\bfcalR}\times \dot{\bfbeta})}{{\cal R}}=\frac{\omega^2 a}{c}\frac{[a\rho\sin^2(\phi-\omega t) +z^2\cos(\phi-\omega t)]\,\hat{\!\bfrho} - \sin(\phi-\omega t)[\rho^2 - \rho a \cos(\phi- \omega t) + z^2]\,\hat{\!\bfphi}}{[\rho^2 + z^2 +a^2-2 \rho a\cos(\phi-\omega t)]^{3/2}}\\
-\frac{\omega^2 a}{c}\frac{z[\rho\cos(\phi - \omega t) -a]\hat{\v z}}{[\rho^2 + z^2 +a^2-2 \rho a\cos(\phi-\omega t)]^{3/2}},\qquad\qquad\qquad\qquad\qquad\qquad\qquad\qquad\\
\frac{\widehat{\bfcalR}\times\dot{\bfbeta}}{{\cal R}}= -\frac{\omega^2 a}{c}\frac{z\sin(\phi-\omega t)\,\hat{\!\bfrho} + z\cos(\phi-\omega t)\,\hat{\!\bfphi} -\rho\sin(\phi-\omega t)\hat{\v z}}{\rho^2 + z^2 +a^2-2 \rho a\cos(\phi-\omega t)},\qquad\qquad\qquad\qquad\qquad\qquad\,\,\,\\
\frac{(\widehat{\bfcalR}\cdot\dot{\bfbeta})\widehat{\bfcalR}\times \bfbeta}{{\cal R}}=\frac{\omega^3 a^2}{c^2}\frac{[\rho\cos(\phi\!-\!\omega t) -a][z\cos(\phi\!-\!\omega t)\,\hat{\!\bfrho}-z\sin(\phi\!-\!\omega t)\,\hat{\!\bfphi} - [\rho\cos(\phi\!-\!\omega t)-a] \hat{\v z}]}{[\rho^2 + z^2 +a^2-2 \rho a\cos(\phi-\omega t)]^{3/2}}.\qquad\nonumber\\
\end{eqnarray}
Now, in cylindrical coordinates we can decompose the fields (C3) and (C4)  in the following form
\begin{eqnarray}
\bfcalE= ({\cal E}^{v}_{\rho} +{\cal E}^{a}_{\rho})\,\hat{\!\bfrho} + ({\cal E}^{v}_{\phi} +{\cal E}^{a}_{\phi})\,\hat{\!\bfphi} + ({\cal E}^{v}_{z} +{\cal E}^{a}_{z})\hat{\v z},\,\,\\
\bfcalB= ({\cal B}^{v}_{\rho} +{\cal B}^{a}_{\rho})\,\hat{\!\bfrho} + ({\cal B}^{v}_{\phi} +{\cal B}^{a}_{\phi})\,\hat{\!\bfphi} + ({\cal B}^{v}_{z} +{\cal B}^{a}_{z})\hat{\v z},
\end{eqnarray}
where ${\cal E}^v_\rho,{\cal E}^v_\phi,{\cal E}^v_z$ and ${\cal B}^v_\rho,{\cal B}^v_\phi,{\cal B}^v_z$ are the components of the velocity fields (those varying as $1/{\cal R}^2$) and ${\cal E}^a_\rho,{\cal E}^a_\phi,{\cal E}^a_z$ and ${\cal B}^a_\rho,{\cal B}^a_\phi,{\cal B}^a_z$ are the components of the acceleration fields (those varying as $1/{\cal R}$). Using (C15)-(C19) evaluated at the retarded time $t_r=t-{\cal R}(t_r)/c$ and (C3) we find the components of the electric field of the encircling dyon
\begin{eqnarray}
{\cal E}^{v}_{\rho}= q\frac{[\rho - a \cos(\phi-\omega t_r)]}{[\rho^2 + z^2 +a^2-2 \rho a\cos(\phi-\omega t_r)]^{3/2}}-g\frac{\omega a}{c}\frac{z\cos(\phi-\omega t_r)}{[\rho^2 + z^2 +a^2-2 \rho a\cos(\phi-\omega t_r)]^{3/2}},\qquad\qquad\,\,\\
{\cal E}^{v}_{\phi}=q\frac{a\sin(\phi-\omega t_r)}{[\rho^2 + z^2 +a^2-2 \rho a\cos(\phi-\omega t_r)]^{3/2}}+g\frac{\omega a}{c}\frac{z\sin(\phi-\omega t_r)}{[\rho^2 + z^2 +a^2-2 \rho a\cos(\phi-\omega t_r)]^{3/2}},\qquad\qquad\,\,\\
{\cal E}^{v}_{z}=q\frac{z}{[\rho^2 + z^2 +a^2-2 \rho a\cos(\phi-\omega t_r)]^{3/2}}+g\frac{\omega a}{c}\frac{[\rho\cos(\phi-\omega t_r)-a]}{[\rho^2 + z^2 +a^2-2 \rho a\cos(\phi-\omega t_r)]^{3/2}},\qquad\qquad\,\,\\
\nonumber {\cal E}^{a}_{\rho}=q\frac{\omega^2a}{c^2}\frac{a\rho\sin^2(\phi-\omega t_r) +z^2\cos(\phi-\omega t_r)}{[\rho^2 + z^2 +a^2-2 \rho a\cos(\phi-\omega t_r)]^{3/2}} + g\frac{\omega^3 a^2}{c^3}\frac{z\cos(\phi-\omega t_r)(\rho\cos(\phi - \omega t_r)-a)}{[\rho^2 + z^2 +a^2-2 \rho a\cos(\phi-\omega t_r)]^{3/2}}\qquad\,\\
- g\frac{\omega^2 a}{c^2}\frac{z\sin(\phi-\omega t_r)}{\rho^2 + z^2 +a^2-2 \rho a\cos(\phi-\omega t_r)},\qquad\qquad\qquad\qquad\qquad\qquad\qquad\qquad\qquad\qquad\quad\,\,\,\,\\
\nonumber{\cal E}^{a}_{\phi}=-q\frac{\omega^2 a}{c^2}\frac{\sin(\phi-\omega t_r)[\rho^2 - \rho a \cos(\phi- \omega t_r) + z^2]}{[\rho^2 + z^2 +a^2-2 \rho a\cos(\phi-\omega t_r)]^{3/2}}-g\frac{\omega^3 a^2}{c^3}\frac{z\sin(\phi-\omega t_r)}{[\rho^2 + z^2 +a^2-2 \rho a\cos(\phi-\omega t_r)]^{3/2}}\\
-g\frac{\omega^2 a}{c^2}\frac{z\cos(\phi-\omega t_r)}{\rho^2 + z^2 +a^2-2 \rho a\cos(\phi-\omega t_r)},\qquad\qquad\qquad\qquad\qquad\qquad\qquad\qquad\qquad\qquad\quad\,\,\\
\nonumber {\cal E}^{a}_{z}=-q\frac{\omega^2 a}{c^2}\frac{z[\rho\cos(\phi - \omega t_r) -a]}{[\rho^2 + z^2 +a^2-2 \rho a\cos(\phi-\omega t_r)]^{3/2}} -g\frac{\omega^3 a^2}{c^3}\frac{[\rho\cos(\phi-\omega t_r)-a]^2}{[\rho^2 + z^2 +a^2-2 \rho a\cos(\phi-\omega t_r)]^{3/2}}\quad\,\,\,\\
+g\frac{\omega^2 a}{c^2}\frac{\rho\sin(\phi-\omega t_r)}{\rho^2 + z^2 +a^2-2 \rho a\cos(\phi-\omega t_r)}.\qquad\qquad\qquad\qquad\qquad\qquad\qquad\qquad\qquad\qquad\quad\,\,\,
\end{eqnarray}
Using the dual changes $q\to g$ and $g\to -q$ in (C22)-(C27) we obtain the corresponding components of the magnetic field of the encircling dyon
\begin{eqnarray}
{\cal B}^{v}_{\rho}= g\frac{[\rho - a \cos(\phi-\omega t_r)]}{[\rho^2 + z^2 +a^2-2 \rho a\cos(\phi-\omega t_r)]^{3/2}}+q\frac{\omega a}{c}\frac{z\cos(\phi-\omega t_r)}{[\rho^2 + z^2 +a^2-2 \rho a\cos(\phi-\omega t_r)]^{3/2}},\qquad\qquad\,\,\\
{\cal B}^{v}_{\phi}=g\frac{a\sin(\phi-\omega t_r)}{[\rho^2 + z^2 +a^2-2 \rho a\cos(\phi-\omega t_r)]^{3/2}}-q\frac{\omega a}{c}\frac{z\sin(\phi-\omega t_r)}{[\rho^2 + z^2 +a^2-2 \rho a\cos(\phi-\omega t_r)]^{3/2}},\qquad\qquad\,\,\\
{\cal B}^{v}_{z}=g\frac{z}{[\rho^2 + z^2 +a^2-2 \rho a\cos(\phi-\omega t_r)]^{3/2}}-q\frac{\omega a}{c}\frac{[\rho\cos(\phi-\omega t_r)-a]}{[\rho^2 + z^2 +a^2-2 \rho a\cos(\phi-\omega t_r)]^{3/2}},\qquad\qquad\,\,\\
\nonumber {\cal B}^{a}_{\rho}=g\frac{\omega^2a}{c^2}\frac{a\rho\sin^2(\phi-\omega t_r) +z^2\cos(\phi-\omega t_r)}{[\rho^2 + z^2 +a^2-2 \rho a\cos(\phi-\omega t_r)]^{3/2}} - q\frac{\omega^3 a^2}{c^3}\frac{z\cos(\phi-\omega t_r)(\rho\cos(\phi - \omega t_r)-a)}{[\rho^2 + z^2 +a^2-2 \rho a\cos(\phi-\omega t_r)]^{3/2}}\qquad\,\\
+ q\frac{\omega^2 a}{c^2}\frac{z\sin(\phi-\omega t_r)}{\rho^2 + z^2 +a^2-2 \rho a\cos(\phi-\omega t_r)},\qquad\qquad\qquad\qquad\qquad\qquad\qquad\qquad\qquad\qquad\quad\,\,\,\,\\
\nonumber{\cal B}^{a}_{\phi}=-g\frac{\omega^2 a}{c^2}\frac{\sin(\phi-\omega t_r)[\rho^2 - \rho a \cos(\phi- \omega t_r) + z^2]}{[\rho^2 + z^2 +a^2-2 \rho a\cos(\phi-\omega t_r)]^{3/2}}+q\frac{\omega^3 a^2}{c^3}\frac{z\sin(\phi\!-\!\omega t_r)}{[\rho^2 + z^2 +a^2-2 \rho a\cos(\phi-\omega t_r)]^{3/2}}\\
+q\frac{\omega^2 a}{c^2}\frac{z\cos(\phi-\omega t_r)}{\rho^2 + z^2 +a^2-2 \rho a\cos(\phi-\omega t_r)},\qquad\qquad\qquad\qquad\qquad\qquad\qquad\qquad\qquad\qquad\quad\,\,\\
\nonumber {\cal B}^{a}_{z}=-g\frac{\omega^2 a}{c^2}\frac{z[\rho\cos(\phi - \omega t_r) -a]}{[\rho^2 + z^2 +a^2-2 \rho a\cos(\phi-\omega t_r)]^{3/2}} +q\frac{\omega^3 a^2}{c^3}\frac{[\rho\cos(\phi-\omega t_r)-a]^2}{[\rho^2 + z^2 +a^2-2 \rho a\cos(\phi-\omega t_r)]^{3/2}}\quad\,\,\,\\
-q\frac{\omega^2 a}{c^2}\frac{\rho\sin(\phi-\omega t_r)}{\rho^2 + z^2 +a^2-2 \rho a\cos(\phi-\omega t_r)}.\qquad\qquad\qquad\qquad\qquad\qquad\qquad\qquad\qquad\qquad\quad\,\,\,
\end{eqnarray}
Equations (C20)-(C33) describe the components of the fields of a non-relativistic dyon in uniform circular motion expressed in terms of the retarded time $t_r=t-{\cal R}(t_r)/c= t-|\v x - \v x_{qg}(t_r)|/c,$ which implicitly depends on the retarded time itself.
\section*{Appendix D. Eliminating the implicit dependence of the retarded time in the fields}
\renewcommand\theequation{D\arabic{equation}}
\setcounter{equation}{0}
In this appendix we will specify the conditions that allow us to eliminate the implicit dependence of the retarded time in the fields of the dyon-solenoid configuration when the dyon is in uniform circular motion, i.e.,  $t_r= t-|\v x - \v x_{qg}(t_r)|/c \to t_r=t-|\v x|/c$.  Since the retarded time in the field components (C22)-(C33) is only present in the functions $\sin(\phi - \omega t_r)$ and $\cos(\phi-\omega t_r)$ then these functions are our quantities of interest. We will see that under certain conditions we can make the replacements $\sin[\phi - \omega (t_r=t-|\v x - \v x_{qg}(t_r)|/c)]\to\sin[\phi - \omega (t_r=t-|\v x|/c)]$ and $\cos[\phi - \omega (t_r=t-|\v x - \v x_{qg}(t_r)|/c)]\to\cos[\phi - \omega (t_r=t-|\v x|/c)]$ and thus eliminate the implicit dependence of the retarded time in the fields.

We now follow an argument similar to that given by Eyges \cite{21}. Using the identities $\sin(A - B)=\sin A\cos B - \cos A \sin B$ and $\cos(A-B)=\sin A\sin B+ \cos A\cos B$ it follows
\begin{eqnarray}
\sin(\phi - \omega t_r)=\sin\phi\cos(\omega t_r) - \cos\phi \sin(\omega t_r),\\
\cos(\phi-\omega t_r)=\sin\phi\sin(\omega t_r)+ \cos\phi\cos(\omega t_r),
\end{eqnarray}
and inserting the retarded time $t_r=t-|\v x - \v x_{qg}(t_r)|/c$ we obtain
\begin{equation}
\sin(\omega t_r)= \sin\bigg[\omega\bigg(t -\frac{|\v x-\v x_{qg}(t_r)|}{c}\bigg) \bigg],\,\,\,\cos(\omega t_r)= \cos\bigg[\omega\bigg(t -\frac{|\v x-\v x_{qg}(t_r)|}{c}\bigg) \bigg].
\end{equation}
The size of the considered distribution is of order $a$ which is the radius of the orbit of the dyon. If $a$ is sufficiently small then we can  make a Taylor expansion in $\v x_{qg}(t_r)$ to the first order in the trigonometric functions in (D3) obtaining
\begin{eqnarray}
\sin(\omega t_r)\approx \sin\bigg[\omega\bigg(t -\frac{|\v x|}{c}\bigg) \bigg] + \v x_{qg}(t_r)\cdot \nabla\sin\bigg[\omega\bigg(t -\frac{|\v x|}{c}\bigg) \bigg],\\
\cos(\omega t_r)\approx \cos\bigg[\omega\bigg(t -\frac{|\v x|}{c}\bigg) \bigg] + \v x_{qg}(t_r)\cdot \nabla\cos\bigg[\omega\bigg(t -\frac{|\v x|}{c}\bigg) \bigg].
\end{eqnarray}
From (C10) it follows  $\v x_{qg}(t_r)=a[\cos(\phi - \omega t_r)\,\hat{\!\bfrho} -\sin(\phi-\omega t_r) \,\hat{\!\bfphi}]$, which is used together with $\nabla = [\partial/\partial\rho]\,\hat{\!\bfrho} + (1/\rho)[\partial/\partial\phi]\,\hat{\!\bfphi} + [\partial/\partial z]\hat{\v z}$ and $|\v x| = \sqrt{\rho^2+z^2}$ to obtain
\begin{eqnarray}
\sin(\omega t_r)\approx \sin\bigg[\omega\bigg(t -\frac{|\v x|}{c}\bigg) \bigg] -\frac{\rho a \omega}{c |\v x|}\cos\bigg[\omega\bigg(t -\frac{|\v x|}{c} \bigg) \bigg]\cos(\omega t_r),\\
\cos(\omega t_r)\approx \cos\bigg[\omega\bigg(t -\frac{|\v x|}{c}\bigg) \bigg] + \frac{\rho a \omega}{c|\v x|}\sin\bigg[\omega\bigg(t -\frac{|\v x|}{c}\bigg) \bigg]\sin(\omega t_r).
\end{eqnarray}
Using (D6) and (D7) in (D1) and (D2) together with the identities $\sin(A - B)=\sin A\cos B - \cos A \sin B$ and $\cos(A-B)=\sin A\sin B+ \cos A\cos B$ we obtain
\begin{eqnarray}
\sin(\phi - \omega t_r)\approx\sin\bigg[\phi-\omega\bigg(t -\frac{|\v x|}{c}\bigg) \bigg]+\frac{\rho \beta}{|\v x|}\cos\bigg[\phi-\omega\bigg(t -\frac{|\v x|}{c}\bigg) \bigg]\cos(\phi-\omega t_r),\\
\cos(\phi - \omega t_r)\approx\cos\bigg[\phi-\omega\bigg(t -\frac{|\v x|}{c}\bigg) \bigg]-\frac{\rho \beta}{|\v x|}\sin\bigg[\phi-\omega\bigg(t -\frac{|\v x|}{c}\bigg) \bigg]\cos(\phi-\omega t_r),
\end{eqnarray}
where $\beta= |\dot{\v x}_{qg}|/c=\omega a/c$ with $|\dot{\v x}_{qg}|= \omega a$ being the magnitude of the velocity of the dyon. The ratio of the second terms on the right-hand side of (D8) and (D9) compared with their corresponding first terms are of order $\beta.$ For a non-relativistic dyon $\beta<<1$ and therefore the second terms on the right-hand side are much smaller compared to the first terms. Accordingly, we can write write the approximation
\begin{eqnarray}
\sin(\phi-\omega t_r)\approx \sin\bigg[\phi-\omega\bigg(t -\frac{|\v x|}{c}\bigg) \bigg],\,\,\,\cos(\phi-\omega t_r)\approx \cos\bigg[\phi-\omega\bigg(t -\frac{|\v x|}{c}\bigg) \bigg].
\end{eqnarray}
Using Eqs.~(D10) and (C20)-(C33) it follows that in the non-relativistic approximation of the fields due to a dyon in uniform circular motion, the retarded time can be expressed as $t_r= t- |\v x|/c$ where now $\v x$ does not depend on the retarded time itself. This property will allow us to integrate the associated fields given in Appendix E.

\renewcommand\theequation{E\arabic{equation}}
\setcounter{equation}{0}
\section*{Appendix E. Proof of $\bfL_{\rm R}=0$ for the dyon-solenoid configuration}
Using the generalised Maxwell equations $\nabla \times \bfcalE=-(1/c)\partial \bfcalB/\partial t - (4\pi/c) \bfcalJ_{m}$ and $\nabla \times \bfcalB= (1/c)\partial \bfcalE/\partial t + (4\pi/c)\bfcalJ_{e}$ we can write Eq.~(48) as
\begin{eqnarray}
\textbf{\bfL}_{\rm R}= \frac{1}{4\pi c^2}\frac{\partial}{\partial t}\int_{V}\v x \times \!\big(\bfcalE\times \v C\big)\,d^3x+\frac{1}{c^2}\int_{V}\v x \times \big(\bfcalJ_{e}\times \v C \big)d^3x-\frac{1}{4\pi c^2}\frac{\partial}{\partial t}\int_{V}\v x \times \!\big(\bfcalB\times \v A\big)\,d^3x-\frac{1}{c^2} \int_{V}\v x \times \big(\bfcalJ_{m}\times \v A \big)d^3x.\,\,\,
\end{eqnarray}
We can now apply (D1) to the dyon-solenoid configuration covering all space expect the dual solenoid. Inserting the position vector $\v x= \rho \hat{\bfrho} + z \hat{\v z},$ the potentials outside the dual solenoid $\v A=\v A_{\rm out}(\v x)=\Phi_{m}\, \hat{\!\bfphi}/(2\pi\rho)$ and $\v C=\v C_{\rm out}(\v x)=-\Phi_{e}\, \hat{\!\bfphi}/(2\pi\rho),$ and $d^3x=\rho d\rho d\phi dz$ in (D1) we obtain
\begin{eqnarray}
\nonumber \textbf{\bfL}_{\rm R}=-\frac{\Phi_{e}}{8\pi^2 c^2}\frac{\partial}{\partial t}\bigg[ \int_{V}\rho\,\hat{\!\bfrho}\times\!\big(\bfcalE\times\,\hat{\!\bfphi}\big)\,d\rho d\phi dz + \int_{V}z\hat{\v z}\times\!\big(\bfcalE\times\,\hat{\!\bfphi}\big)\, d\rho d\phi dz \bigg] \,\,\,\\
-\frac{\Phi_{e}}{2\pi c^2}\bigg[\int_{V} \rho\,\hat{\!\bfrho}\times\!\big(\bfcalJ_{e}\times\,\hat{\!\bfphi}\big)\, d\rho d\phi dz+\int_{V}z\hat{\v z}\times\!\big(\bfcalJ_{e}\times\,\hat{\!\bfphi}\big)\, d\rho d\phi dz \bigg]\nonumber\quad\\
\nonumber - \frac{\Phi_{m}}{8\pi^2 c^2}\frac{\partial}{\partial t} \bigg[\int_{V}\rho\,\hat{\!\bfrho}\times\!\big(\bfcalB\times\,\hat{\!\bfphi}\big)\,d\rho d\phi dz+\int_{V}z\hat{\v z}\times\!\big(\bfcalB\times\,\hat{\!\bfphi}\big)\,d\rho d\phi dz\bigg]\,\\
- \frac{\Phi_{m}}{2\pi c^2}\bigg[\int_{V} \rho\,\hat{\!\bfrho}\times\!\big(\bfcalJ_{m}\times\,\hat{\!\bfphi}\big)\, d\rho d\phi dz + \int_{V}z\hat{\v z}\times\!\big(\bfcalJ_{m}\times\,\hat{\!\bfphi}\big)\, d\rho d\phi dz \bigg].
\end{eqnarray}
The fields are Li\'enard-Wiechert fields produced by the encircling dyon which from (C20) and (C21) can be decomposed as $\bfcalE= ({\cal E}^{v}_{\rho} +{\cal E}^{a}_{\rho})\,\hat{\!\bfrho} + ({\cal E}^{v}_{\phi} +{\cal E}^{a}_{\phi})\,\hat{\!\bfphi} + ({\cal E}^{v}_{z} +{\cal E}^{a}_{z})\hat{\v z}$ and $\bfcalB= ({\cal B}^{v}_{\rho} +{\cal B}^{a}_{\rho})\,\hat{\!\bfrho} + ({\cal B}^{v}_{\phi} +{\cal B}^{a}_{\phi})\,\hat{\!\bfphi} + ({\cal B}^{v}_{z} +{\cal B}^{a}_{z})\hat{\v z}$ where ${\cal E}^v_\rho,{\cal E}^v_\phi,{\cal E}^v_z$ and ${\cal B}^v_\rho,{\cal B}^v_\phi,{\cal B}^v_z$ are the components of the velocity fields and ${\cal E}^a_\rho,{\cal E}^a_\phi,{\cal E}^a_z$ and ${\cal B}^a_\rho,{\cal B}^a_\phi,{\cal B}^a_z$ are the components of the acceleration fields . The currents of the encircling dyon can be written as $\bfcalJ_{e}=({\cal J}_e)_\rho\,\hat{\!\bfrho}+({\cal J}_e)_\phi\,\hat{\!\bfphi}+({\cal J}_e)_z\hat{\v z}$ and $\bfcalJ_{m}=({\cal J}_m)_\rho\,\hat{\!\bfrho}+({\cal J}_m)_\phi\,\hat{\!\bfphi}+({\cal J}_m)_z\hat{\v z}$. Using these results in (E2) and performing the specified operations, we obtain
\begin{eqnarray}
\textbf{\bfL}_{\rm R}=\frac{\Phi_{e}\,\hat{\!\bfphi}}{8\pi^2 c^2}\bigg[\frac{\partial I_1}{\partial t} + \frac{\partial I_2}{\partial t}+\frac{\partial I_3}{\partial t} + \frac{\partial I_4}{\partial t}\bigg]+\frac{\Phi_{e}\,\hat{\!\bfphi}}{2\pi c^2}[I_5 + I_{6}]+ \frac{\Phi_{m}\,\hat{\!\bfphi}}{8\pi^2 c^2}\bigg[\frac{\partial I_7}{\partial t} + \frac{\partial I_8}{\partial t}+\frac{\partial I_9}{\partial t} + \frac{\partial I_{10}}{\partial t}\bigg]+ \frac{\Phi_{m}\,\hat{\!\bfphi}}{2\pi c^2}[I_{11} + I_{12}], 
\end{eqnarray}
where the integrals $I_{1}-I_{12}$ are defined as
\begin{eqnarray}
I_1= \int_{V}\rho {\cal E}^{v}_{\rho}\,d\rho d\phi dz,\,\, I_2=\int_{V}\rho {\cal E}^{a}_{\rho}\,d\rho d\phi dz,\,\, I_3= \int_{V}z {\cal E}^{v}_{z}\, d\rho d\phi dz,\,\,\,I_4= \int_{V}z {\cal E}^{a}_{z}\, d\rho d\phi dz,\,\,\,\,\,\\
I_5= \int_{V}\rho({\cal J}_{e})_{\rho}\,d\rho d\phi dz,\,\, I_{6}= \int_{V}z ({\cal J}_{e})_{z}\, d\rho d\phi dz,\qquad\qquad\qquad\qquad\qquad\qquad\qquad\qquad\,\,\,\,\,\\
I_7= \int_{V}\rho {\cal B}^{v}_{\rho}\,d\rho d\phi dz,\,\, I_8=\int_{V}\rho {\cal B}^{a}_{\rho}d\rho d\phi dz,\,\, I_9= \int_{V}z {\cal B}^{v}_{z}\, d\rho d\phi dz,\,\,I_{10}= \int_{V}z {\cal B}^{a}_{z}\, d\rho d\phi dz,\,\,\,\,\\
I_{11}=\int_{V}\rho ({\cal J}_{m})_{\rho}\,d\rho d\phi dz,\quad I_{12}= \int_{V}z ({\cal J}_{m})_{z}\, d\rho d\phi dz,\qquad\qquad\qquad\qquad\qquad\qquad\qquad\,\,\,\,\,\,\,
\end{eqnarray}
The field components ${\cal E}^v_\rho,{\cal E}^v_\phi,{\cal E}^v_z,{\cal E}^a_\rho,{\cal E}^a_\phi,{\cal E}^a_z$ and ${\cal B}^v_\rho,{\cal B}^v_\phi,{\cal B}^v_z,{\cal B}^a_\rho,{\cal B}^a_\phi,{\cal B}^a_z$ as well as the current components $({\cal J}_e)_\rho,({\cal J}_e)_\phi,({\cal J}_e)_z$ and $({\cal J}_m)_\rho,({\cal J}_m)_\phi,({\cal J}_m)_z$ should now be specified. For simplicity, we consider the fields and currents due to a dyon encircling the dual solenoid in non-relativistic uniform circular motion along the $x$-$y$ plane. More specifically, we assume that the dyon is moving with constant angular velocity $\omega$ in a circle of fixed radius $a$ around the dual solenoid. Field coordinates are denoted as $\rho,\phi,z$ and the source coordinates specifying the position of the dyon are denoted as $a,\omega t,0.$ In this case we can use the field components in (C23)-(C33) together with the approximation in (D10). To find the components of the current densities we use $\bfcalJ_{e}= q  \dot{\v x}_{qg}\delta[\v x- \v x_{qg}(t)]$ and $\bfcalJ_{m}= g  \dot{\v x}_{qg}\delta[\v x- \v x_{qg}(t)]$ together with (C11) and $\delta[\v x- \v x_{qg}(t)]=\delta(\rho - a)\delta(\phi - \omega t)\delta(z)/\rho$ to obtain $\bfcalJ_{e}=[q\omega a/\rho][\sin(\phi - \omega t)\,\hat{\!\bfrho} +\cos(\phi-\omega t) \,\hat{\!\bfphi}]\delta(\rho - a)\delta(\phi - \omega t)\delta(z)$ and $\bfcalJ_{m}=[g\omega a/\rho][\sin(\phi - \omega t)\,\hat{\!\bfrho} +\cos(\phi-\omega t) \,\hat{\!\bfphi}]\delta(\rho - a)\delta(\phi - \omega t)\delta(z)$. Therefore
\begin{eqnarray}
({\cal J}_{e})_{\rho}=\frac{q \omega a}{\rho} \sin(\phi - \omega t)\delta(\rho - a)\delta(\phi - \omega t)\delta(z),\,\,\,\,\,\\
({\cal J}_{e})_{\phi}=\frac{q \omega a}{\rho} \cos(\phi - \omega t)\delta(\rho - a)\delta(\phi - \omega t)\delta(z),\,\,\,\,\\
({\cal J}_{e})_{z}=0,\qquad\qquad\qquad\qquad\qquad\qquad\qquad\quad\quad\,\,\,\,\\
({\cal J}_{m})_{\rho}=\frac{g \omega a}{\rho} \sin(\phi - \omega t)\delta(\rho - a)\delta(\phi - \omega t)\delta(z),\,\,\,\,\,\\
({\cal J}_{m})_{\phi}=\frac{g \omega a}{\rho} \cos(\phi - \omega t)\delta(\rho - a)\delta(\phi - \omega t)\delta(z),\,\,\,\,\\
({\cal J}_{m})_{z}=0.\qquad\qquad\qquad\qquad\qquad\qquad\qquad\qquad\quad
\end{eqnarray}
Let us now proceed to evaluate the components within the brackets in (E3).
\vskip 2pt
\noindent \textbf{(i) Proof of} $\bm {\partial I_1/\partial t=0}$. Using (C22) and $I_1$ in (E4) it follows
\begin{eqnarray}
\nonumber I_1= q\int^{\infty}_{R}\rho d\rho \int^{+\infty}_{-\infty} dz \oint_{C}\frac{[\rho - a \cos(\phi-\omega (t- \sqrt{\rho^2 + z^2}/c))]\,d\phi}{[\rho^2 + z^2 +a^2-2 \rho a\cos(\phi-\omega(t- \sqrt{\rho^2 + z^2}/c))]^{3/2}}\qquad\qquad\,\,\,\\
-g\frac{\omega a}{c}\int^{\infty}_{R}\rho d\rho \oint_{C}d\phi \int^{+\infty}_{-\infty}\frac{z\cos(\phi-\omega (t- \sqrt{\rho^2 + z^2}/c))dz}{[\rho^2 + z^2 +a^2-2 \rho a\cos(\phi-\omega (t- \sqrt{\rho^2 + z^2}/c))]^{3/2}}.\qquad
\end{eqnarray}
We observe that the integrand in the second term on the right-hand side is an odd function of $z$. Therefore it directly follows
\begin{equation}
\int^{+\infty}_{-\infty}\frac{z\cos(\phi-\omega (t- \sqrt{\rho^2 + z^2}/c))dz}{[\rho^2 + z^2 +a^2-2 \rho a\cos(\phi-\omega (t- \sqrt{\rho^2 + z^2}/c))]^{3/2}}=0,
\end{equation}
which is used in (E14) to obtain
\begin{eqnarray}
I_1= q\int^{\infty}_{R}\rho d\rho \int^{+\infty}_{-\infty} dz \oint_{C}\frac{[\rho - a \cos(\phi-\omega (t- \sqrt{\rho^2 + z^2}/c))]\,d\phi}{[\rho^2 + z^2 +a^2-2 \rho a\cos(\phi-\omega(t- \sqrt{\rho^2 + z^2}/c))]^{3/2}}.
\end{eqnarray}
Inserting the operator $\partial/\partial t$ in (E16) we obtain
\begin{eqnarray}
\frac{\partial I_1}{\partial t}= q\int^{\infty}_{R}\rho^2 d\rho \int^{+\infty}_{-\infty} dz \oint_{C}\frac{\partial}{\partial t}\bigg[\frac{[\rho - a \cos(\phi-\omega (t- \sqrt{\rho^2 + z^2}/c))]}{[\rho^2 + z^2 +a^2-2 \rho a\cos(\phi-\omega(t- \sqrt{\rho^2 + z^2}/c))]^{3/2}}\bigg]\,d\phi.
\end{eqnarray}
The partial derivative gives
\begin{eqnarray}
\nonumber \frac{\partial}{\partial t}\bigg[\frac{[\rho - a \cos(\phi-\omega (t- \sqrt{\rho^2 + z^2}/c))]}{[\rho^2 + z^2 +a^2-2 \rho a\cos(\phi-\omega(t- \sqrt{\rho^2 + z^2}/c))]^{3/2}}\bigg]\qquad\qquad\qquad\qquad\qquad\qquad\qquad\\
\nonumber = - \frac{a\omega \sin(\phi- \omega(t-\sqrt{\rho^2 + z^2}/c))}{[\rho^2 + z^2 +a^2-2 \rho a\cos(\phi-\omega(t- \sqrt{\rho^2 + z^2}/c))]^{3/2}}\qquad\qquad\qquad\qquad\quad \,\,\,\,\,\,\,\\
\nonumber +\frac{3a\rho^2\omega \sin(\phi-\omega(t- \sqrt{\rho^2 + z^2}/c))}{[\rho^2 + z^2 +a^2-2 \rho a\cos(\phi-\omega(t- \sqrt{\rho^2 + z^2}/c))]^{5/2}}\qquad\qquad\qquad\qquad\qquad\\
-\frac{3a^2\rho\omega\sin(\phi-\omega (t- \sqrt{\rho^2 + z^2}/c))\cos(\phi-\omega (t- \sqrt{\rho^2 + z^2}/c))}{[\rho^2 + z^2 +a^2-2 \rho a\cos(\phi-\omega(t- \sqrt{\rho^2 + z^2}/c))]^{5/2}},\qquad\quad\qquad
\end{eqnarray}
which is used in (E17) to obtain
\begin{eqnarray}
\nonumber \frac{\partial I_1}{\partial t}= -qa \omega \int^{\infty}_{R}\rho d\rho \int^{+\infty}_{-\infty} dz \oint_{C}\frac{\sin(\phi- \omega(t-\sqrt{\rho^2 + z^2}/c))d\phi}{[\rho^2 + z^2 +a^2-2 \rho a\cos(\phi-\omega(t- \sqrt{\rho^2 + z^2}/c))]^{3/2}}\qquad\qquad\quad\,\,\,\,\\
\nonumber +3 q a\omega  \int^{\infty}_{R}\rho^3 d\rho \int^{+\infty}_{-\infty} dz \oint_{C}\frac{\sin(\phi-\omega(t- \sqrt{\rho^2 + z^2}/c))d\phi}{[\rho^2 + z^2 +a^2-2 \rho a\cos(\phi-\omega(t- \sqrt{\rho^2 + z^2}/c))]^{5/2}}\qquad\qquad\,\,\,\,\\
-3 q a^2\omega  \int^{\infty}_{R}\rho^2 d\rho \int^{+\infty}_{-\infty}\!\! dz \oint_{C}\frac{\sin(\phi\!-\!\omega(t\!-\! \sqrt{\rho^2 \!+\! z^2}/c))\cos(\phi\!-\!\omega (t\!-\! \sqrt{\rho^2 \!+\! z^2}/c))d\phi}{[\rho^2 + z^2 +a^2-2 \rho a\cos(\phi-\omega(t- \sqrt{\rho^2 + z^2}/c))]^{5/2}}.\quad\,\,\,\,\,\,\,\,\,\,\,\,
\end{eqnarray}
The azimuthal integrals can be evaluated via a Contour integration. Here we use \textit{Mathematica} to calculate them
\begin{eqnarray}
\int^{2\pi n}_{0}\frac{\sin(\phi- \omega(t-\sqrt{\rho^2 + z^2}/c))d\phi}{[\rho^2 + z^2 +a^2-2 \rho a\cos(\phi-\omega(t- \sqrt{\rho^2 + z^2}/c))]^{3/2}}=0,\qquad\qquad\\
\int^{2\pi n}_{0}\frac{\sin(\phi- \omega(t-\sqrt{\rho^2 + z^2}/c))d\phi}{[\rho^2 + z^2 +a^2-2 \rho a\cos(\phi-\omega(t- \sqrt{\rho^2 + z^2}/c))]^{5/2}}=0,\qquad\qquad\\
\int^{2\pi n}_{0}\frac{\sin(\phi- \omega(t-\sqrt{\rho^2 + z^2}/c))\cos(\phi- \omega(t-\sqrt{\rho^2 + z^2}/c))d\phi}{[\rho^2 + z^2 +a^2-2 \rho a\cos(\phi-\omega(t- \sqrt{\rho^2 + z^2}/c))]^{5/2}}=0,\quad
\end{eqnarray}
where $n$ is the winding number. Using (E20)-(E22) in (E19) we obtain
\begin{equation}
\frac{\partial I_1}{\partial t}=0.
\end{equation}
\vskip 2pt
\noindent \textbf{(ii) Proof of} $\bm {\partial I_2/\partial t=0}$. Using (C25) and $I_2$ in (E4) we obtain
\begin{eqnarray}
\nonumber I_2= q\frac{\omega^2a^2}{c^2} \int^{\infty}_{R}\rho^2 d\rho \int^{+\infty}_{-\infty} dz \oint_{C} \frac{\sin^2(\phi-\omega(t-\sqrt{\rho^2 + z^2}/c))d\phi}{[\rho^2 + z^2 +a^2-2 \rho a\cos(\phi-\omega(t-\sqrt{\rho^2 + z^2}/c))]^{3/2}}\qquad\\
\nonumber +q\frac{\omega^2a}{c^2} \int^{\infty}_{R}\rho d\rho \int^{+\infty}_{-\infty}z^2 dz \oint_{C} \frac{\cos(\phi-\omega(t-\sqrt{\rho^2 + z^2}/c))d\phi}{[\rho^2 + z^2 +a^2-2 \rho a\cos(\phi-\omega(t-\sqrt{\rho^2 + z^2}/c))]^{3/2}}\\
\nonumber + g\frac{\omega^3 a^2}{c^3}\int^{\infty}_{R}\rho^2 d\rho\oint_{C}d\phi \int^{+\infty}_{-\infty}  \frac{z\cos^2(\phi-\omega(t-\sqrt{\rho^2 + z^2}/c))dz}{[\rho^2 + z^2 +a^2-2 \rho a\cos(\phi-\omega(t-\sqrt{\rho^2 + z^2}/c))]^{3/2}}\\
\nonumber -g\frac{\omega^3 a^3}{c^3}\int^{\infty}_{R}\rho d\rho  \oint_{C}d\phi\int^{+\infty}_{-\infty}\frac{z \cos(\phi-\omega(t-\sqrt{\rho^2 + z^2}/c))dz}{[\rho^2 + z^2 +a^2-2 \rho a\cos(\phi-\omega(t-\sqrt{\rho^2 + z^2}/c))]^{3/2}}\\
- g\frac{\omega^2 a}{c^2}\int^{\infty}_{R}\rho d\rho \oint_{C}d\phi\int^{+\infty}_{-\infty}  \frac{z\sin(\phi-\omega(t-\sqrt{\rho^2 + z^2}/c))dz}{\rho^2 + z^2 +a^2-2 \rho a\cos(\phi-\omega(t-\sqrt{\rho^2 + z^2}/c))}.\qquad
\end{eqnarray}
Using (E15) the fourth term on the right-hand side of (D24) vanishes. We observe that the integrad in the fifth term of (24) is an odd function of $z$and therefore it directly follows
\begin{equation}
\int^{+\infty}_{-\infty}\frac{z\sin(\phi-\omega(t-\sqrt{\rho^2 + z^2}/c))d\phi}{\rho^2 + z^2 +a^2-2 \rho a\cos(\phi-\omega(t-\sqrt{\rho^2 + z^2}/c))}=0.
\end{equation}
On the other hand, the integrand in the second term of the right-hand side of (E24) is also an odd function of $z$ and therefore
\begin{equation}
\int^{+\infty}_{-\infty}\frac{z\cos^2(\phi-\omega(t-\sqrt{\rho^2 + z^2}/c))dz}{[\rho^2 + z^2 +a^2-2 \rho a\cos(\phi-\omega(t-\sqrt{\rho^2 + z^2}/c))]^{3/2}}=0.
\end{equation}
Using (E15), (E25) and (E26) in (E24) we obtain
\begin{eqnarray}
\nonumber I_2= q\frac{\omega^2a^2}{c^2} \int^{\infty}_{R}\rho^2 d\rho \int^{+\infty}_{-\infty} dz \oint_{C} \frac{\sin^2(\phi-\omega(t-\sqrt{\rho^2 + z^2}/c))d\phi}{[\rho^2 + z^2 +a^2-2 \rho a\cos(\phi-\omega(t-\sqrt{\rho^2 + z^2}/c))]^{3/2}}\qquad\\
+q\frac{\omega^2a}{c^2} \int^{\infty}_{R}\rho d\rho \int^{+\infty}_{-\infty}z^2 dz \oint_{C} \frac{\cos(\phi-\omega(t-\sqrt{\rho^2 + z^2}/c))d\phi}{[\rho^2 + z^2 +a^2-2 \rho a\cos(\phi-\omega(t-\sqrt{\rho^2 + z^2}/c))]^{3/2}}.
\end{eqnarray}
Inserting the operator $\partial/\partial t$ in (E27) we obtain
\begin{eqnarray}
\nonumber \frac{\partial I_2}{\partial t}= q\frac{\omega^2a^2}{c^2} \int^{\infty}_{R}\rho^2 d\rho \int^{+\infty}_{-\infty} dz \oint_{C} \frac{\partial}{\partial t}\bigg[\frac{\sin^2(\phi-\omega(t-\sqrt{\rho^2 + z^2}/c))}{[\rho^2 + z^2 +a^2-2 \rho a\cos(\phi-\omega(t-\sqrt{\rho^2 + z^2}/c))]^{3/2}}\bigg]d\phi\qquad\\
+q\frac{\omega^2a}{c^2} \!\int^{\infty}_{R}\rho d\rho \!\int^{+\infty}_{-\infty}\!\!z^2 dz \!\oint_{C} \frac{\partial}{\partial t}\bigg[\frac{\cos(\phi-\omega(t-\sqrt{\rho^2 + z^2}/c))}{[\rho^2 \!+\! z^2 \!+\!a^2\!-\!2 \rho a\cos(\phi-\omega(t-\sqrt{\rho^2 + z^2}/c))]^{3/2}}\bigg]d\phi.\quad\,\,\,\,\,\,
\end{eqnarray}
The partial derivatives in (E28) give
\begin{eqnarray}
\nonumber \frac{\partial}{\partial t}\bigg[\frac{\sin^2(\phi-\omega(t-\sqrt{\rho^2 + z^2}/c))}{[\rho^2 + z^2 +a^2-2 \rho a\cos(\phi-\omega(t-\sqrt{\rho^2 + z^2}/c))]^{3/2}}\bigg]\qquad\qquad\qquad\qquad\qquad\qquad\qquad\,\,\,\,\\
\nonumber = -\frac{2\omega\sin(\phi-\omega(t-\sqrt{\rho^2 + z^2}/c))\cos(\phi-\omega(t-\sqrt{\rho^2 + z^2}/c))}{[\rho^2 + z^2 +a^2-2 \rho a\cos(\phi-\omega(t-\sqrt{\rho^2 + z^2}/c))]^{3/2}}\qquad\qquad\,\,\,\,\\
+\frac{3 a \rho \sin^3(\phi-\omega(t-\sqrt{\rho^2 + z^2}/c))}{[\rho^2 + z^2 +a^2-2 \rho a\cos(\phi-\omega(t-\sqrt{\rho^2 + z^2}/c))]^{5/2}},\qquad\qquad\qquad\quad\,\,\,\\
\nonumber \frac{\partial}{\partial t}\bigg[\frac{\cos(\phi-\omega(t-\sqrt{\rho^2 + z^2}/c))}{[\rho^2 \!+\! z^2 \!+\!a^2\!-\!2 \rho a\cos(\phi-\omega(t-\sqrt{\rho^2 + z^2}/c))]^{3/2}}\bigg]\qquad\qquad\qquad\qquad\qquad\qquad\qquad\qquad\\
\nonumber =\frac{\omega\sin(\phi-\omega(t-\sqrt{\rho^2 + z^2}/c))}{[\rho^2 \!+\! z^2 \!+\!a^2\!-\!2 \rho a\cos(\phi-\omega(t-\sqrt{\rho^2 + z^2}/c))]^{3/2}}\qquad\qquad\qquad\qquad\qquad\\
+ \frac{3a\rho\omega \sin(\phi-\omega(t-\sqrt{\rho^2 + z^2}/c))\cos(\phi-\omega(t-\sqrt{\rho^2 + z^2}/c))}{[\rho^2 \!+\! z^2 \!+\!a^2\!-\!2 \rho a\cos(\phi-\omega(t-\sqrt{\rho^2 + z^2}/c))]^{5/2}},\qquad\quad\,\,\,
\end{eqnarray}
which are used in (E28) to obtain
\begin{eqnarray}
\nonumber \frac{\partial I_2}{\partial t}= -2q\omega\frac{\omega^2a^2}{c^2} \!\int^{\infty}_{R}\rho^2\! d\rho \!\int^{+\infty}_{-\infty}\!\! dz \oint_{C} \frac{\sin(\phi\!-\!\omega(t\!-\!\sqrt{\rho^2 \!+\! z^2}/c))\cos(\phi\!-\!\omega(t\!-\!\sqrt{\rho^2 \!+\! z^2}/c))d\phi}{[\rho^2 + z^2 +a^2-2 \rho a\cos(\phi-\omega(t-\sqrt{\rho^2 + z^2}/c))]^{3/2}}\qquad\\
\nonumber +q\frac{3\omega^2a^3}{c^2} \int^{\infty}_{R}\rho^3 d\rho \int^{+\infty}_{-\infty} dz \oint_{C}\frac{\sin^3(\phi-\omega(t-\sqrt{\rho^2 + z^2}/c))d\phi}{[\rho^2 + z^2 +a^2-2 \rho a\cos(\phi-\omega(t-\sqrt{\rho^2 + z^2}/c))]^{5/2}}\qquad\,\,\,\\
\nonumber +q\frac{\omega^3a}{c^2} \!\int^{\infty}_{R}\rho d\rho \!\int^{+\infty}_{-\infty}\!\!z^2 dz \!\oint_{C} \frac{\sin(\phi-\omega(t-\sqrt{\rho^2 + z^2}/c))d\phi}{[\rho^2 \!+\! z^2 \!+\!a^2\!-\!2 \rho a\cos(\phi-\omega(t-\sqrt{\rho^2 + z^2}/c))]^{3/2}}\qquad\qquad\,\,\,\\
+3q\frac{\omega^3a^2}{c^2} \!\!\int^{\infty}_{R}\rho^2 d\rho \!\int^{+\infty}_{-\infty}\!\!z^2 dz \!\oint_{C}\frac{\sin(\phi\!-\!\omega(t\!-\!\sqrt{\rho^2 \!+\! z^2}/c))\cos(\phi\!-\!\omega(t\!-\!\sqrt{\rho^2 \!+\! z^2}/c))d\phi}{[\rho^2 \!+\! z^2 \!+\!a^2\!-\!2 \rho a\cos(\phi-\omega(t-\sqrt{\rho^2 + z^2}/c))]^{5/2}}.\quad\,\,\,
\end{eqnarray}
The third term on the right-hand side of (E31) vanishes on account of (E20) and the fourth term on the right-hand side vanishes on account of (E22). The remaining azimuthal integrals may be evaluated via a Contour integration. Here we use \textit{Mathematica} to calculate them
\begin{eqnarray}
\int^{2\pi n}_{0} \frac{\sin(\phi\!-\!\omega(t\!-\!\sqrt{\rho^2 \!+\! z^2}/c))\cos(\phi\!-\!\omega(t\!-\!\sqrt{\rho^2 \!+\! z^2}/c))d\phi}{[\rho^2 + z^2 +a^2-2 \rho a\cos(\phi-\omega(t-\sqrt{\rho^2 + z^2}/c))]^{3/2}}=0,\\
\int^{2\pi n}_{0}\frac{\sin^3(\phi-\omega(t-\sqrt{\rho^2 + z^2}/c))d\phi}{[\rho^2 + z^2 +a^2-2 \rho a\cos(\phi-\omega(t-\sqrt{\rho^2 + z^2}/c))]^{5/2}}=0.
\end{eqnarray}
Using (E20), (E22) and (E32)-(E33) in (E31) we obtain
\begin{equation}
\frac{\partial I_2}{\partial t}=0.
\end{equation}
\vskip 2pt
\noindent \textbf{(iii) Proof of} $\bm {\partial I_3/\partial t=0}$. Using (C24) and $I_3$ in (E4) we obtain
\begin{eqnarray}
\nonumber I_3 = q\int^{\infty}_{R} d\rho \int^{+\infty}_{-\infty}z^2 dz \oint_{C}\frac{d\phi}{[\rho^2 + z^2 +a^2-2 \rho a\cos(\phi-\omega (t-\sqrt{\rho^2 + z^2}/c))]^{3/2}}\qquad\\
+g\frac{\omega a}{c} \int^{\infty}_{R}d\rho \oint_{C}d\phi \int^{+\infty}_{-\infty} \frac{z[\rho\cos(\phi-\omega (t-\sqrt{\rho^2 + z^2}/c))-a]dz}{[\rho^2 + z^2 +a^2-2 \rho a\cos(\phi-\omega (t-\sqrt{\rho^2 + z^2}/c))]^{3/2}}.
\end{eqnarray}
The integrand in the second term of the right-hand side is an odd function of $z$ and therefore
\begin{equation}
\int^{+\infty}_{-\infty} \frac{z[\rho\cos(\phi-\omega (t-\sqrt{\rho^2 + z^2}/c))-a]dz}{[\rho^2 + z^2 +a^2-2 \rho a\cos(\phi-\omega (t-\sqrt{\rho^2 + z^2}/c))]^{3/2}}=0.
\end{equation}
Using (E36) in (E35) we obtain
\begin{equation}
I_3 = q\int^{\infty}_{R} d\rho \int^{+\infty}_{-\infty}z^2 dz \oint_{C}\frac{d\phi}{[\rho^2 + z^2 +a^2-2 \rho a\cos(\phi-\omega (t-\sqrt{\rho^2 + z^2}/c))]^{3/2}}.
\end{equation}
Inserting the operator $\partial/\partial t$ in (E37) we obtain
\begin{equation}
\frac{\partial I_3}{\partial t}= q\int^{\infty}_{R} d\rho \int^{+\infty}_{-\infty}z^2 dz \oint_{C}\frac{\partial}{\partial t}\bigg[\frac{1}{[\rho^2 + z^2 +a^2-2 \rho a\cos(\phi-\omega (t-\sqrt{\rho^2 + z^2}/c))]^{3/2}}\bigg]d\phi.
\end{equation}
The partial derivative gives
\begin{eqnarray}
\nonumber \frac{\partial}{\partial t}\bigg[\frac{1}{[\rho^2 + z^2 +a^2-2 \rho a\cos(\phi-\omega (t-\sqrt{\rho^2 + z^2}/c))]^{3/2}}\bigg]\qquad\qquad\qquad\\
=\frac{3 a \rho \omega \sin(\phi-\omega (t-\sqrt{\rho^2 + z^2}/c))}{[\rho^2 + z^2 +a^2-2 \rho a\cos(\phi-\omega (t-\sqrt{\rho^2 + z^2}/c))]^{5/2}},
\end{eqnarray}
which is used in (E38) to obtain
\begin{equation}
\frac{\partial I_3}{\partial t}= 3qa\omega\int^{\infty}_{R}\rho d\rho \int^{+\infty}_{-\infty}z^2 dz \oint_{C} \frac{\sin(\phi-\omega (t-\sqrt{\rho^2 + z^2}/c))d\phi}{[\rho^2 + z^2 +a^2-2 \rho a\cos(\phi-\omega (t-\sqrt{\rho^2 + z^2}/c))]^{5/2}}.
\end{equation}
The azimuthal integral vanishes on account of (E21) and therefore we obtain
\begin{equation}
\frac{\partial I_3}{\partial t}=0.
\end{equation}
\noindent \textbf{(iv) Proof of} $\bm {\partial I_4/\partial t=0}$. Using (C27) and $I_4$ in (E4) we obtain
\begin{eqnarray}
\nonumber I_4=-q\frac{\omega^2 a}{c^2}\int^{\infty}_{R} d\rho \int^{+\infty}_{-\infty}z^2 dz \oint_{C} \frac{[\rho\cos(\phi - \omega (t-\sqrt{\rho^2 + z^2}/c)) -a]d\phi}{[\rho^2 + z^2 +a^2-2 \rho a\cos(\phi-\omega (t-\sqrt{\rho^2 + z^2}/c))]^{3/2}} \\
\nonumber -g\frac{\omega^3 a^2}{c^3}\int^{\infty}_{R} d\rho \oint_{C}d \phi\int^{+\infty}_{-\infty} \frac{z[\rho\cos(\phi-\omega(t-\sqrt{\rho^2 + z^2}/c)))-a]^2 dz}{[\rho^2 + z^2 +a^2-2 \rho a\cos(\phi-\omega(t-\sqrt{\rho^2 + z^2}/c) )]^{3/2}}\\
+g\frac{\omega^2 a}{c^2}\int^{\infty}_{R}\rho d\rho \oint_{C}d\phi\int^{+\infty}_{-\infty}  \frac{z\sin(\phi-\omega (t-\sqrt{\rho^2 + z^2}/c))dz}{\rho^2 + z^2 +a^2-2 \rho a\cos(\phi-\omega (t-\sqrt{\rho^2 + z^2}/c))},\,\,\,\,
\end{eqnarray}
The third term on the right-hand side of (E42) vanishes on account of (E25). The integrand in the second term is an odd function of $z$ and therefore
\begin{equation}
\int^{+\infty}_{-\infty}\frac{z[\rho\cos(\phi-\omega(t-\sqrt{\rho^2 + z^2}/c)))-a]^2 dz}{[\rho^2 + z^2 +a^2-2 \rho a\cos(\phi-\omega(t-\sqrt{\rho^2 + z^2}/c) )]^{3/2}}=0,
\end{equation}
which is used in (E42) together with (E25) to obtain
\begin{eqnarray}
I_4=-q\frac{\omega^2 a}{c^2}\int^{\infty}_{R} d\rho \int^{+\infty}_{-\infty}z^2 dz \oint_{C} \frac{[\rho\cos(\phi - \omega (t-\sqrt{\rho^2 + z^2}/c)) -a]d\phi}{[\rho^2 + z^2 +a^2-2 \rho a\cos(\phi-\omega (t-\sqrt{\rho^2 + z^2}/c))]^{3/2}}.
\end{eqnarray}
Inserting the operator $\partial/\partial t$ in (E44) we obtain
\begin{equation}
\frac{\partial I_4}{\partial t}= -q\frac{\omega^2 a}{c^2}\int^{\infty}_{R} d\rho \int^{+\infty}_{-\infty}z^2 dz \oint_{C}\frac{\partial}{\partial t} \bigg[\frac{[\rho\cos(\phi - \omega (t-\sqrt{\rho^2 + z^2}/c)) -a]}{[\rho^2 + z^2 +a^2-2 \rho a\cos(\phi-\omega (t-\sqrt{\rho^2 + z^2}/c))]^{3/2}}\bigg]d\phi.\qquad
\end{equation}
The partial derivative gives
\begin{eqnarray}
\nonumber \frac{\partial}{\partial t} \bigg[\frac{[\rho\cos(\phi - \omega (t-\sqrt{\rho^2 + z^2}/c)) -a]}{[\rho^2 + z^2 +a^2-2 \rho a\cos(\phi-\omega (t-\sqrt{\rho^2 + z^2}/c))]^{3/2}}\bigg]\qquad\qquad\qquad\qquad\qquad\qquad\qquad\\
\nonumber =\frac{\omega \rho \sin(\phi - \omega (t-\sqrt{\rho^2 + z^2}/c))}{[\rho^2 + z^2 +a^2-2 \rho a\cos(\phi-\omega (t-\sqrt{\rho^2 + z^2}/c))]^{3/2}}\qquad\qquad\qquad\qquad\\
\nonumber +\frac{3a\omega\rho^2\sin(\phi - \omega (t-\sqrt{\rho^2 + z^2}/c))\cos(\phi - \omega (t-\sqrt{\rho^2 + z^2}/c))}{[\rho^2 + z^2 +a^2-2 \rho a\cos(\phi-\omega (t-\sqrt{\rho^2 + z^2}/c))]^{5/2}}\qquad\,\,\,\\
-\frac{3a^2\omega\rho\sin(\phi - \omega (t-\sqrt{\rho^2 + z^2}/c))}{[\rho^2 + z^2 +a^2-2 \rho a\cos(\phi-\omega (t-\sqrt{\rho^2 + z^2}/c))]^{5/2}},\qquad\qquad\qquad\,\,\,\,\,\,\,
\end{eqnarray}
which is used in (E45) to obtain
\begin{eqnarray}
\nonumber \frac{\partial I_4}{\partial t}= -q\frac{\omega^3 a}{c^2}\int^{\infty}_{R}\rho d\rho \int^{+\infty}_{-\infty}z^2 dz \oint_{C}\frac{\sin(\phi - \omega (t-\sqrt{\rho^2 + z^2}/c))d\phi}{[\rho^2 + z^2 +a^2-2 \rho a\cos(\phi-\omega (t-\sqrt{\rho^2 + z^2}/c))]^{3/2}}\qquad\qquad\\
\nonumber -3q\frac{\omega^3 a^2}{c^2}\int^{\infty}_{R}\rho^2 d\rho \int^{+\infty}_{-\infty}z^2 dz \oint_{C}\frac{\sin(\phi - \omega (t-\sqrt{\rho^2 + z^2}/c))\cos(\phi - \omega (t-\sqrt{\rho^2 + z^2}/c))d\phi}{[\rho^2 + z^2 +a^2-2 \rho a\cos(\phi-\omega (t-\sqrt{\rho^2 + z^2}/c))]^{5/2}}\\
+3q\frac{\omega^3 a^3}{c^2}\int^{\infty}_{R}\rho d\rho \int^{+\infty}_{-\infty}z^2 dz \oint_{C}\frac{\sin(\phi - \omega (t-\sqrt{\rho^2 + z^2}/c))d\phi}{[\rho^2 + z^2 +a^2-2 \rho a\cos(\phi-\omega (t-\sqrt{\rho^2 + z^2}/c))]^{5/2}}.\qquad\,\,\,\,
\end{eqnarray}
The three azimuthal integrals in the right-hand side of (E47) vanish on account of (E20)-(E22) and therefore we obtain
\begin{equation}
\frac{\partial I_4}{\partial t}=0.
\end{equation}
\vskip 2pt
\noindent \textbf{(v) Proof of} $\bm {I_5=0}$. Using (E9) and $I_5$ in (E5) we obtain
\begin{equation}
I_5=q \omega a \int^{\infty}_{R}\delta(\rho - a) d\rho\int^{+\infty}_{-\infty}\delta(z)dz \oint_{C} \sin(\phi - \omega t)\delta(\phi - \omega t)d\phi.
\end{equation}
The first two integrals give
\begin{equation}
\int^{\infty}_{R}\delta(\rho - a) d\rho=\Theta(a-R)=1\,\,\,(a>R),\quad\int^{+\infty}_{-\infty}\delta(z)dz=1,
\end{equation}
which are used in (E49) to obtain
\begin{equation}
I_5=q \omega a\oint_{C} \sin(\phi - \omega t)\delta(\phi - \omega t)d\phi.
\end{equation}
Using \textit{Mathematica} the azimuthal integral gives
\begin{equation}
\int^{2\pi n}_{0}\sin(\phi - \omega t)\delta(\phi - \omega t)d\phi=0.
\end{equation}
Using (E52) in (E51) we obtain
\begin{equation}
I_5=0.
\end{equation}
\vskip 2pt
\noindent \textbf{(vi) Proof of} $\bm {I_6=0}$. From (E10) we have $({\cal J}_{e})_{z}=0$ which is used in $I_6$ in (E5) to obtain
\begin{equation}
I_6=0.
\end{equation}
\vskip 2pt
\noindent \textbf{(vii)-(xii) Proofs of} $\bm {\partial I_7/\partial t=0, \partial I_8/\partial t=0, \partial I_9/\partial t=0, \partial I_{10}/\partial t=0, I_{11}=0,$} \textbf{and} $\bm {I_{12}=0.}$ The integrals $I_{7}-I_{12}$ are of the same form as those in $I_1-I_6.$ In fact by making the dual changes $q\to g$ and $g\to -q$ in the integrals $I_1-I_6$ we obtain the integrals $I_{7}-I_{12}$. Therefore
\begin{equation}
\frac{\partial I_7}{\partial t}=0,\,\,\,\frac{\partial I_8}{\partial t}=0,\,\,\,\frac{\partial I_9}{\partial t}=0,\,\,\,\frac{\partial I_{10}}{\partial t}=0,\,\,\, I_{11}=0, \,\,\,I_{12}=0.
\end{equation}
Using (E23), (E34), (E41), (E48), and (E53)-(E54) in (E3) we finally obtain
\begin{equation}
\textbf{\bfL}_{\rm R}=0.
\end{equation}
Since this relation is valid for a dyon with charges $q\not=0$ and $g\not=0$ and for a dual solenoid with fluxes $\Phi_e\not=0$  and $\Phi_m\not =0$ then it is valid for the particular case in which the dyon has the charges $q=0$  and $g\not=0$ (a magnetic monopole) and the dual solenoid has the fluxes $\Phi_m=0$ and $\Phi_e\not=0$ (an electric solenoid):
\begin{equation}
\textbf{\bfL}_{\rm R}(q=0, \Phi_m=0)=0,
\end{equation}
which corresponds to the monopole-solenoid configuration (see (20)).

\end{document}